\documentclass[fleqn,usenatbib]{mnras}

\usepackage{newtxtext,newtxmath}
 
\usepackage[T1]{fontenc}

\DeclareRobustCommand{\VAN}[3]{#2}
\let\VANthebibliography\thebibliography
\def\thebibliography{\DeclareRobustCommand{\VAN}[3]{##3}\VANthebibliography}

\usepackage{graphicx}	
\usepackage{amsmath}	
\usepackage{amssymb}	
\usepackage{subfigure}
\usepackage{paralist}

\everymath{\displaystyle}

\newcommand{\todo}[1]{{\textcolor{black}{#1}}}
\newcommand\tE{t_{\rm E}}
\newcommand\RE{R_{\rm E}}{}
\newcommand\Dl{D_{\rm l}}
\newcommand\Ds{D_{\rm s}}

\makeatletter
\newcommand*{\smallrel}[2][.8]{%
  \mathrel{\mathpalette{\smallrel@{#1}}{#2}}%
}
\newcommand*{\smallrel@}[3]{%
  \sbox0{$#2\vcenter{}$}%
  \dimen@=\ht0 %
  \raise\dimen@\hbox{%
    \scalebox{#1}{%
      \raise-\dimen@\hbox{$#2#3\m@th$}%
    }%
  }%
}
\makeatother

\usepackage{pdflscape}
\usepackage{hhline}
\usepackage{setspace}
\usepackage[flushleft]{threeparttable}

\title[VVV Microlensing]{A microlensing search of 700 million VVV light curves}

\author[A. Husseiniova et al.]{
Andrea Husseiniova,$^{1,2}$
Peter McGill,$^{1}$\thanks{E-mail: pm625@cam.ac.uk (PM), nwe@ast.cam.ac.uk (NWE)}
Leigh C. Smith$^{1}$
and N. Wyn Evans$^{1}$
\\
$^{1}$Institute of Astronomy, University of Cambridge, Madingley Rd, Cambridge CB3 0HA, UK\\
$^{2}$School of Physics and Astronomy, University of Edinburgh, James Clerk Maxwell Building,
Peter Guthrie Tait Road, Edinburgh EH9 3FD, UK
}

\date{Accepted XXX. Received YYY; in original form ZZZ}

\pubyear{2021}

\begin{document}
\label{firstpage}
\pagerange{\pageref{firstpage}--\pageref{lastpage}}
\maketitle

\begin{abstract}
The VISTA Variables in the Vıa Lactea (VVV) survey and its extension, have been monitoring about $560$ square degrees of sky centred on the Galactic bulge and inner disc for nearly a decade. The photometric catalogue contains of order 10$^9$ sources monitored in the $K_s$ band down to $18$ mag over hundreds of epochs from 2010-2019. Using these data we develop a decision tree classifier to identify microlensing events. As inputs to the tree, we extract a few physically motivated features as well as simple statistics ensuring a good fit to a microlensing model both on and off the event amplification. This produces a fast and efficient classifier trained on a set of simulated microlensing events and catacylsmic variables, together with flat baseline light curves randomly chosen from the VVV data. The classifier achieves 97 per cent accuracy in identifying simulated microlensing events in a validation set. We run the classifier over the VVV data set and then visually inspect the results, which produces a catalogue of 1\,959 microlensing events. For these events, we provide the Einstein radius crossing time via a Bayesian analysis. \todo{The spatial dependence on recovery efficiency of our classifier is well characterised, and this} allows us to compute spatially resolved completeness maps as a function of Einstein crossing time over the VVV footprint. We compare our approach to previous microlensing searches of the VVV. We highlight the importance of Bayesian fitting to determine the microlensing parameters for events with surveys like VVV with sparse data.
\end{abstract}




\section{Introduction}

Historically, microlensing projects were amongst the first to acquire large photometric datasets and systematically search through them for very rare events~\cite[e.g.][]{Au93,Al00,Ca05}.  The probability that a star is microlensed -- or the all-sky source-averaged optical depth to microlensing -- is very low $\sim10^{-7}$ \citep[e.g.][]{Be02}, and so the events are much less numerous than stellar variability.

As a star passes behind a lensing object, its light gradually becomes brighter. Its brightness peaks when the star is at its closest approach to the lensing object, then fades again as it moves away. The shape of the light curve is symmetric, achromatic and does not repeat, at least in the simplest instance that both source and lens are point-like. The light curve for such a standard microlensing event has a simple form first derived by \citet{Pa86}. Specifically, assuming a linear trajectory between the point source and point lens (PSPL hereafter), the amplification is
\begin{equation}\label{eq:PSPL_model}
    A(t) = \frac{u^2 + 2}{u\sqrt{u^2+4}}, \quad{}
    u(t) = \sqrt{u^2_0 + \left(\frac{t-t_0}{\tE}\right)^2}.
\end{equation}
Here, $t_0$ is the time of the event peak, $u_0$ is the impact parameter (normalised by the angular Einstein radius $\RE/\Dl$, where $\RE$ is the Einstein Radius and $\Dl$ is the distance from the observer to the lens) and $\tE$ is the Einstein crossing time. \todo{In the case of unresolved blended light from another source the observed amplification of the blend is,} 
\begin{equation}
   \todo{A_{\text{obs}}(t) = f_{bl} A(t) + \left(1 - f_{bl}\right)}.
\end{equation}
\todo{Here, $f_{bl}$ is the fraction of lensed source flux to the total blend flux.} Of the parameters that can be extracted from the standard PSPL light curve, only $\tE$ contains physical information, as it is related to the mass of the lens $M$ and the relative source lens transverse velocity $v$ by
\begin{equation}
\tE= \frac{\RE}{v},\qquad \RE = \left (\frac{4 GM}{c^2} \frac{\Dl (\Ds-\Dl)}{\Ds}\right)^{1/2}.
\end{equation}
Here, $\Ds$ is the distance from observer to source. This degeneracy means that physical interpretation of the event is ambiguous in the absence of further data. Some microlensing events deviate from the standard form, or can be predicted ahead of time, enabling the degeneracies to be wholly or partially lifted \citep[e.g.,][]{Go92,Go94,Ben98,Ev03, McGill2019a}, though these exotica can be hard to pick out. In this paper, we direct our attention exclusively to standard PSPL events. As microlensing traces the mass of the Galaxy rather than the light, it provides a unique and complementary probe to other surveys of the Milky Way. Given substantial numbers of PSPL events, properties of the source and lens populations can be inferred, providing the selection function of the survey is well characterized. This has led to new insights into the structure of the Galaxy, the Magellanic Clouds and M31~\citep[e.g.,][]{Ki94,Ev03,PH03,Ca11,Wyrzykowski2011,Wyrzykowski2011b,We16}.

Bulge microlensing surveys in the optical hark back to the very beginnings of the subject~\citep{Ud92,Al95}. The high number of background stars provides ample sources, whilst the density of disc and bulge stars means that the optical depth to microlensing is high at $\sim 10^{-6}$.  The OGLE or Optical Gravitational Lensing Experiment ~\citep{Ud97,Ud15,Wy15} has heroically maintained a decades long campaign to monitor events towards the centre of the Galaxy in optical passbands at high cadence. The most recent results from the OGLE-IV campaign provide measurements of the optical depth and event rate across the Galactic bulge and disc, based on  nearly $9000$ events~\citep{Mr19,Mr20new}. However, there are parts of the bulge that are heavily obscured by extinction, hampering searches in optical passbands. It was early realised that photometry in the near-IR overcomes this obstacle and can provide for many more locations throughout the bulge to search for events~\citep{Go95}. 

The VISTA Variables in the Vıa Lactea Survey \citep[VVV,][]{Mi10} is a near-IR variability Survey that scans 560 square degrees in the inner Milky Way using the Visible and Infrared Survey Telescope for Astronomy (VISTA). \citet{Na17} quickly realised the importance of this dataset for microlensing in the very central parts of the Galaxy. They reported the discovery of 182 new microlensing events, based on observations acquired between 2010 and 2015, in the three innermost tiles of the survey. This covers an area $1.68^\circ \ge \ell \ge -2.68^\circ$ and $0.65^\circ \ge b\ge -0.46^\circ$.  Later, \citet{NavarroLatitude,NavarroForesaken} extended the microlensing search to cover 14 tiles, encompassing all of the Galactic plane in VVV. They presented a catalogue of the 630 microlensing events, covering the region within $10.44^\circ \ge \ell \ge -10^\circ$ and $0.65^\circ \ge b\ge -0.46^\circ$. Even so, this is only 14 of the 348 tiles in the entire VVV survey. Searching through the whole VVV dataset remains an intimidating task.

In order to search large surveys, automatic classification of microlensing events is required. An early attempt was made by \citet{Bel03} using neural networks, but the subject has come of age in recent years. First, \citet{Pr14} drew attention to the usefulness of the von Neumann statistic, which is the mean square of successive differences divided by the variance of data points in the light curve. This is very fast to compute, especially compared to the process of fitting a PSPL light curve. They found it gave excellent performance in identifying simulated microlensing events injected into real light curves from the Palomar Transient Factory~\citep{Of12}, which have irregular temporal sampling. Then, \cite{Wy15} used a random forest classifier to extract microlensing events for OGLE-III, finding 3\,700 unique events from over 150 million light curves in data taken between 2001-2009. After preliminary cuts removing obvious variable stars and other contaminants that did not possess single-humped light curves, \cite{Wy15} were left with 48\,112 light curves for further classification. They then used a Random Forest classifier for the last stage. This took as input 27 features extracted from the primarily $I$ band, but also some $V$ band, light curves. This final step extracted the 3\,700 unique events -- including 1\,409 events that had not been detected before in real-time by the OGLE Early Warning System. This was the first study to harness the power of machine learning algorithms to find rare microlensing phenomena and to derive the detection efficiency. 
\citet{Go19} again used a Random Forest classifier, but
now with 47 statistical features extracted from light curves. They included variable stars, cataclysmic variables as well as constant light curves in their training set, so that the machine learning algorithm is tasked with all the work and there is no preliminary filtering. They trained the classifier using the OGLE-II microlensing dataset, and tested it on
Palomar and Zwicky Transient Factory survey data.
Finally, we note that the problem of automatic identification of microlensing events in which the system is not PSPL -- sometimes showing substantial deviations from the standard \citet{Pa86} curve -- has also began to receive attention \citep{Mr2020,Zh20,Khakpash2021}.

The paper is arranged as follows. In Section~\ref{sec:data}, we provide an overview of the characteristics of the data, together with details of the simulated light curves that are used to populate the training set and build completeness maps. Section~\ref{sec:search} describes the construction of the classifier, the choice of the features used to train and tune it, as well as its performance on the validation data. The results obtained when applied to the VVV data, together with the completeness maps, are given in Section~\ref{sec:res}. The catalogue of 1\,959 microlensing events, together with their fitted parameters, is presented in Appendix A. Finally, Section~\ref{sec:comp} compares the properties of our sample of microlensing events with the earlier work of \citet{Navarro2018, NavarroLatitude, NavarroForesaken, NavarroFardisk}, and a brief outline of the inference on $t_{E}$ for our sample of events is presented in Section \ref{sec:tE_inference}.

\section{Data}
\label{sec:data}

\subsection{Observed Light Curves}

The VVV survey, and its temporal and spatial extension the VVV eXtended (VVVX; \citealt{vvvx}) survey, sample $>10^9$ sources in the southern Galactic disc and bulge over hundreds of epochs across the 2010-2019 calendar years. These surveys utilised the VISTA Infrared Camera \citep[VIRCAM;][]{vircam} and cover a total area of approximately 1\,700 square degrees, although only the 560 square degrees of the VVV survey contains data covering the full time baseline. The value in these surveys for the identification and study of Galactic microlensing events is well documented \citep{Na17,2019MNRAS.487L...7M,NavarroLatitude,NavarroForesaken,NavarroFardisk}.

Production of version two of the VVV Infrared Astrometric Catalogue (see \citealt{virac} for details of version one) has yielded a time series photometry database for of order $10^9$ stars in the VVV area (approx. 560 square degrees) using VVV and VVVX data. Hereafter we will refer to the VVV survey plus the VVVX survey data covering the VVV area as simply the VVV survey. A detailed description of the database and data processing is outside the scope of this paper. To briefly summarise, VIRCAM stacked pawprint images from the VVV survey were processed with a modified version of the DoPHOT profile fit image reduction software \citep{dophot1, dophot2}. Photometric measurements were coarsely calibrated using a globally optimized zero point plus illumination map model. A further fine calibration stage was developed and deployed to reduce high spatial frequency coherent structure in the maps of photometric residuals of individual observations to the survey average, which reduces scatter in the light curves. In tandem with the latter calibration stage we also scale the pipeline photometric uncertainties to better account for residual scatter across each observation. The photometric calibration stages are described in detail in Smith et al. (in prep).

\subsection{Simulated Light Curves}
\label{sec:simulations}

In order to evaluate an algorithm for discrimination between microlensing and non-microlensing, we adopt a simulation and data-driven approach. This requires large samples of microlensing, non-microlensing variable, and non-variable events.

A sample of non-variable events is straightforward to produce, as a random selection of light curves contains only a small fraction of sources with significant real variability in VVV data. The production of sufficiently large samples of variable events requires simulation. \citet{Go19} identified cataclysmic variables (CVs), RR Lyrae and Cepheid variables as likely contaminants in searches for microlensing events. Due to the long baseline of our data and hence good phase coverage, we don't anticipate significant contamination from RR Lyrae or Cepheid variable stars in our search; there is however potential for contamination by CVs due to the irregular observing pattern and large gaps in coverage between observing seasons. We therefore restrict our simulation of non-microlensing variables to CVs only.

It is important that our simulated events adopt realistic observing cadence, measurement uncertainty, outlier incidence and scale, spatial and brightness distributions. These are necessary for an accurate evaluation of both the performance of our selection criteria and the overall completeness, which we expect to vary as a function of sky position and brightness due to the variable nature of the underlying survey.

The most straightforward approach to meeting the above requirements is \todo{to perform a catalog-level simulation \citep[e.g.][]{Mr19,Wy15, Wyrzykowski2009}}. Specifically, we take a random selection of real light curves and inject signals into them. This naturally adopts the true observing cadence, spatial and brightness distributions, and with some care outlying data points can also be preserved. The response of the brightness measurement uncertainty to the signal amplification is the only remaining detail to be addressed. To deal with this, we first select $10^8$ detections at random and reject the $\sim{}14$\% of these that are either obvious blends or non- $K_s$ observations. In the range $K_s = [10,18)$ with a step size of $0.1$~mag, we select from this detection pool those which are within $0.1$~mag of each interval and measure the integer percentiles of their photometric uncertainties in the range $[0,100]$. This produces a photometric uncertainty lookup table of $101$ percentile measurements at $80$ $K_s$ band magnitude points, from which we can select appropriate photometric uncertainties at a given magnitude and relative observation quality.

The process of simulating a single light curve begins with a selection of a random real light curve as a `seed'. The seed light curve sets the observing sequence, and its median dictates the baseline magnitude $m_0$. The magnitude at a given time $m_t$ is then drawn from the appropriate model and random parameter distribution (see Sections \ref{pspl_model} and \ref{cv_model}).

An important detail is that the observing sequence is not only the epochs of the observations but also their quality. To assess the quality of each real observation from the seed light curve, we find the percentile, $P_t$, corresponding to the measured magnitude uncertainty for the measured magnitude from our photometric uncertainty lookup table using linear interpolation. Finally, for each observation with a model magnitude, $m_t$, we select an appropriate magnitude error $\sigma_{m_t}$ through linear interpolation of our photometric uncertainty lookup table as a function of $m_t$ and $P_t$ and scatter the model $m_t$ by a magnitude drawn from ${\mathcal N}(0,\sigma_{m_t})$.

\begin{figure}
  \begin{center}
    \includegraphics[width=\linewidth,keepaspectratio]{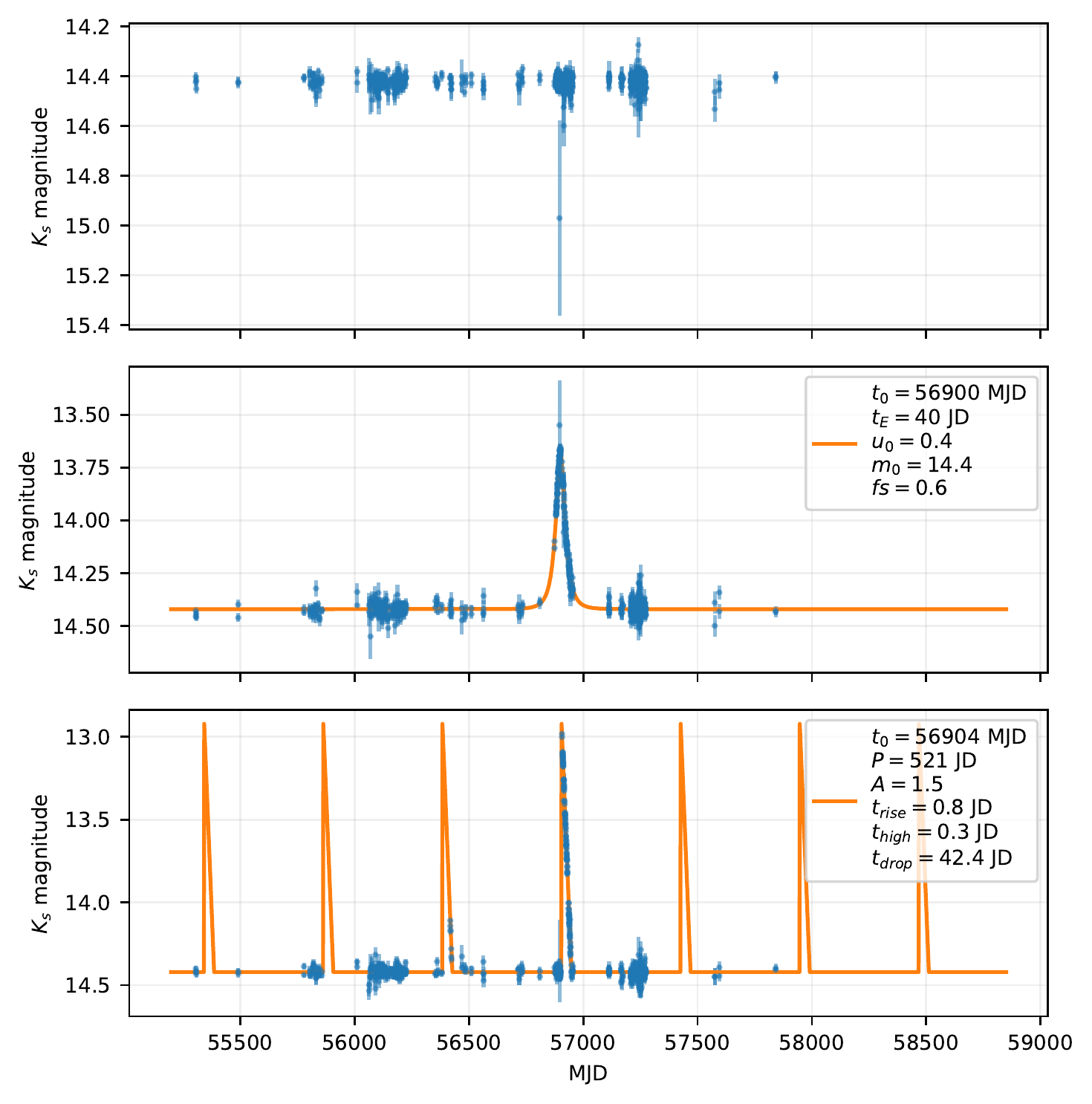}
    \caption{\textbf{Top}:~The seed light curve used for the simulations in the lower panels. \textbf{Middle}:~A simulated PSPL light curve and model. \textbf{Bottom}:~A simulated CV light curve and model. All model parameters were hand-selected from the simulation distributions to best highlight the response of the photometric uncertainties to the variability.}
    \label{LC_sim_example}
  \end{center}
\end{figure}

The unique identifier of the seed light curve is preserved throughout this process so that we may later recover location information. Fig.~\ref{LC_sim_example} shows a representative seed light curve and examples of simulated microlensing and CV light curves using this seed.

\subsubsection{Point source point lens simulation model}\label{pspl_model}

To simulate a microlensing event, we randomly assign the remaining PSPL model parameters by drawing from uniform distributions (\todo{$\mathcal{U}$(lower limit, upper limit)}) \todo{on a linear scale} as follows: $t_0 \sim {\mathcal U}(2009,2021)$~Julian~year, $t_{\rm E} \sim {\mathcal U}(10~{\rm minutes}, 1000~{\rm days})$, $u_0 \sim {\mathcal U}(10^{-3},5)$, $\todo{f_{bl}} \sim {\mathcal U}(0,1)$. The microlensing light curve model magnitudes for each seed observation epoch and the PSPL parameters are generated using equation~\ref{eq:PSPL_model}.

\subsubsection{CV simulation model}\label{cv_model}

To simulate a CV light curve, we adopt the same approach as \citet{Go19} but with slightly tweaked parameter distributions.
\todo{The original (optical) data source that inspired the choices is \citet{Ro76}.} To summarise, CV light curves are characterised by periodic outbursts with a relatively sharp rise in flux to $m_{\rm min}$ (where $m_{\rm min}=m_0-A$ and $A$ is the outburst amplitude), a short period of high flux at $m_{\rm min}$ and a slower drop in flux back to the baseline level of $m_0$. For each simulated light curve, we randomly select parameters as follows: Period, $P$, from ${\mathcal U}(100,200)$~days, truncated on the low end at $30$~days; Duration, $d$, from ${\mathcal U}(3,0.1P)$~days; Amplitude, $A$, from ${\mathcal U}(0.5,5.0)$~mag; Rise time, $t_{\rm rise}$, from ${\mathcal U}(0.5,1.0)$~days; Time at high flux, $t_{\rm high}$, from ${\mathcal U}(0.4,0.2)$~days, truncated on the low end at $0$~days; Drop time, $t_{\rm drop}$, is simply $t_{\rm drop}=d-t_{\rm rise}-t_{\rm high}$.

The truncation on the period and $t_{\rm high}$ parameter distributions are not mentioned by \citet{Go19}, but are necessary purely to ensure times are always positive and that the gradient of the flux drop is always negative. CVs with periods shorter than this are likely to be reasonably well sampled in phase space by our long baseline data and therefore not cause significant contamination anyway. The gradient of the rising portion of the outburst is therefore $g_{\rm rise}=-A/t_{\rm rise}$~mag~day$^{-1}$, and the gradient of the dropping portion is $g_{\rm drop}=A/t_{\rm drop}$~mag~day$^{-1}$.

The magnitude of a simulated CV light curve at time since the start of a given outburst, $t'$, is:
\begin{equation} \label{eq:CV_model}
       M(t) = 
\begin{cases}
      m_0\!+\!t'g_{\rm rise}                                &\mbox{if }  t'\le{}t_{\rm rise} \\
m_0\!-\!A                                             &\mbox{if }  t_{\rm rise}<t'\le{}(t_{\rm rise}\!+\!t_{\rm high}) \\
m_0\!-\!A\!+\!g_{\rm drop}(t'\!-\!t_{\rm rise}\!-\!t_{\rm high}) &\mbox{if } (t_{\rm rise}\!+\!t_{\rm high})<t'\le{}d
\end{cases}
\end{equation}

\section{Search Algorithm}
\label{sec:search}

In our search algorithm, we opt for a feature extraction based approach to distinguish microlensing events from other light curves~\citep[c.f.][]{Pr14,Wy15,Go19}. However, instead of computing many general summary statistics of the light curve, we focus on a few interpretable features based on microlensing model fitting and checking.

\subsection{Maximum likelihood estimates}
\label{sec:mle}

This first feature captures how well a microlensing model fits the data compared to a constant magnitude model. A full Bayesian analysis of this problem obtaining posterior samples via Markov chain Monte Carlo (MCMC) methods for both the microlensing and constant light curve models for all $\sim10^9$ sources is computationally unfeasible. Even optimized gradient-based Hamiltonian MCMC samplers\footnote{e.g. \url{https://caustic.readthedocs.io/}}, typically take $\sim 10$s to converge on light curves with hundreds of data points \citep[e.g.][]{Go20}. Therefore, we are restricted to use maximum likelihood fits, which are much cheaper to compute (typically $ 10^{-1}$~s per source for PSPL microlensing models).   

Let a data point be the set of time, magnitude, and measurement error $D_{i} = \{t_{i},m_{i}, \sigma_{m_{i}}\}$, and a light curve be the set of $n$ data points $\mathcal{D} = \{D_{i}\}_{i=1}^{n}$. Let $m_{\mathcal{M}}(t_{i};\boldsymbol{\theta})$ be the predicted magnitude at time $t_{i}$ for model $\mathcal{M}$ with parameters $\boldsymbol{\theta}$. Under the assumption that we know $t_{i}$ exactly and the magnitudes are corrupted with independent Gaussian noise with variance $ \sigma^{2}_{m_{i}}$, the log likelihood of the light curve given the model for a set of parameter values is,
\begin{equation}
    \ln p(\mathcal{D}|\boldsymbol{\theta},\mathcal{M}) = -\frac{1}{2}\boldsymbol{r}^{T}\textbf{\textsf{K}}^{-1}\boldsymbol{r}-\frac{1}{2}|\textbf{\textsf{K}}| - \frac{n}{2}\ln2\pi
    \label{eq:log_like}
\end{equation}
Here $\boldsymbol{r} = \left[m_{1}-m_{\mathcal{M}}(t_{1};\boldsymbol{\theta}),...,m_{n}-m_{\mathcal{M}}(t_{n};\boldsymbol{\theta})\right]^T$ is the residual vector between the model and data of length $n$, and $\textbf{\textsf{K}}=\text{diag}(\sigma^{2}_{m_{1}},..,\sigma^{2}_{m_{n}})$ is an $n\times n$ diagonal covariance matrix. The maximum likelihood estimate (MLE) for a given model $
\boldsymbol{\theta}_{\text{mle}}$ is obtained by maximising equation~\ref{eq:log_like} (or minimizing $-\ln p(\mathcal{D}|\boldsymbol{\theta},\mathcal{M})$). The corresponding maximum likelihood value is  $\ln p (\mathcal{D}|\boldsymbol{\theta}_\text{mle},\mathcal{M}).$ 

The first model fitted is a constant magnitude model, specifically, $m_{\text{const}}(t;C)=C$ where $\boldsymbol{\theta}\equiv C$. In this case, the maximum log likelihood solution can be obtained analytically and cheaply. Let $\textbf{1}$ be a vector of ones $[1,..,1]^{T}$ with length $n$, and $\textbf{m}$ be the vector of observed magnitudes from the light curve $[m_{1},..,m_{n}]^{T}$. Then, the maximum likelihood solution for the constant magnitude model is,
\begin{equation}
    C_{\text{mle}} = \left[\boldsymbol{1}^{T}\textbf{\textsf{K}}^{-1}\boldsymbol{1}\right]^{-1}\boldsymbol{1}^{T}\textbf{\textsf{K}}^{-1}\boldsymbol{m}=\frac{\sum_{i=1}^{n}m_{i}\sigma^{-2}_{m_{i}}}{\sum^{n}_{i=1}\sigma^{-2}_{m_{i}}}.
\end{equation}
This is the inverse variance weighted average of the light curve magnitudes.

The second model is the standard \cite{Paczynski1996} PSPL model, already introduced in equation~\ref{eq:PSPL_model}. Assuming some non-zero amount of blended light not from source, then the PSPL model is,
\begin{equation}
    m_{\text{PSPL}}(t;\boldsymbol{\theta}) = m_{0} - 2.5\log_{10}[f_{\rm bl}A(t)+1-f_{\rm bl}].
\end{equation}
Here, $m_{0}$ is source baseline magnitude, and $\boldsymbol{\theta}\equiv[t_0,t_{\rm E},u_0,m_0,f_{\rm bl}]$.
Unlike the constant model, the maximum likelihood solution for the PSPL model has to be obtained numerically. We used the differential evolution optimization algorithm \citep{st97} implemented by the Scipy Python package \citep{2020SciPy-NMeth} to find the MLE.

Differential evolution is a non-gradient based global optimization algorithm. It works by stochastically evolving a population of candidate solutions in the parameter space according to a chosen strategy. We used the "best1bin" strategy, which picks and assesses trial solutions from the population of candidate solutions using the binomial distribution. We chose a population size of $15$, a maximum number of iterations of $1\,000$, a convergence tolerance of $0.01$, differential weight of $0.5$, and a recombination constant of $0.7$. Differential evolution also requires the parameters are bounded. The bounds (lower, upper) used for $t_{\rm E},u_0,f_{\rm bl}$ were fixed across all sources at ($1.0$, $10^3$) days, (10$^{-3}$, $3.0$), and ($0.0$, $1.0$) respectively. The bounds for $t_{0}$ and $m_{0}$ were set dynamically for each source to between the minimum and maximum values of observation epoch and magnitude in the light curve, respectively. 

We note that it is possible to compute gradients of the likelihood with respect to the model parameters in the PSPL model, which permits the use of gradient based optimizers \citep[e.g. the Broyden–Fletcher–Goldfarb–Shanno (BFGS) algorithm,][]{Br76}. Although these are typically faster than differential evolution because they require substantially less likelihood evaluations, we found differential evolution reliably found the global solution, whereas the gradient base optimizers often got stuck in local minima. This was likely due to the optimization problem being difficult because of the highly non-linear PSPL model and typically sparse and noisy VVV data. Despite differential evolution being comparatively slow, it is still feasible to run the optimization for all sources in the VVV survey. The PSPL maximum likelihood optimization took $253$~ms per light curve per \todo{CPU} core on AMD EPYC $7452$ chips. It was run on a total of $669\,244\,443$ light curves in the VVV on $1024$ \todo{CPU} cores, totaling $47\,033$ CPU hours of computation.

\subsection{Model comparison}

With the MLE of both the constant and PSPL models in hand for every source, we can asses the relative quality of their fits to the data. We choose the Akaike information criterion \citep[AIC,][]{ak81} to compare the models. AIC is an estimator of the in-sample predictive performance of a model on a data set and is defined as,
\begin{equation}
    \text{AIC}_{\mathcal{M}} = 2k_{\mathcal{M}} - 2\ln p(\mathcal{D}|\boldsymbol{\theta}_{\text{mle}}, \mathcal{M}),
    \label{eq:AIC}
\end{equation}
where $k_{\mathcal{M}}$ is the number of parameters being estimated in the model $\mathcal{M}$. AIC is the log predictive density given the MLE with a bias correction term to account for how much fitting a model of k parameters will increase the predictive accuracy alone \citep{ge14}. The underlying assumption that permits adding the number of model parameters is that the posterior is a multivariate Gaussian distribution. This is generally not true for the PSPL model, meaning equation~\ref{eq:AIC} is an approximation. The main reason for using AIC is that it is readily calculable from the MLE. We are unable to calculate more accurate measures of model predictive performance, for example the Deviance information criteria \citep{sp02} or cross-validations scores \citep[e.g.][]{ve17}, as they require samples from the posterior distribution which are too expensive to compute. The difference in AIC scores can then be computed for the PSPL and constant models as
\begin{equation}
 \Delta\text{AIC} = \text{AIC}_{\text{const}} - \text{AIC}_{\text{PSPL}},
\end{equation}
where a positive number indicates the PSPL model better explains the data, approximately accounting for the increased complexity of the PSPL model.

\subsection{Model checking}

The $\Delta\text{AIC}$ score only assesses the relative model performance between the PSPL and constant models. Therefore, we are still blind to cases where both models have poor fits to the data. To remedy this, we compute quality cuts and features based on the PSPL MLE solution. The cuts are applied to all light curves, both real and simulated. This serves two purposes. In the case of simulated microlensing events, it ensures the simulations produce events that were reasonably able to be detected by the survey. In the case of real data, the cuts eliminate likely non-microlensing events or events not well detected in the survey. Features are used to train the classifier to extract microlensing events.

\subsubsection{Quality cuts}
\label{sec:quality}

The first quality cut aims to ensure that the microlensing events are well covered by the data. This eliminates events for which the PSPL fit is being driven by only a few data points on the amplification. Moreover, it aims to ensure we extract microlensing events with sufficient coverage that they can have useful inferences drawn about the PSPL parameters. Using the PSPL maximum likelihood parameters for each source, we require the number of data points before and after two $t_{\rm E}$ of $t_{0}$ to be greater than ten ($n_{\text{before}}>10$ and $n_{\text{after}}>10$). We also require more than five data points within one $t_{\rm E}$ of $t_{0}$ ($n_{\text{amp}}>5$).


Next we make two cuts that ensure that the microlensing amplification is significant compared to noise in the baseline. For both of these cuts, we define points on the baseline to be at least two $t_{\rm E}$ away from $t_{0}$ and points on the amplification to be within two $t_{\rm E}$ of $t_{0}$. First, we require at least three data points on the amplification above three standard deviations of the \todo{mean} of the data points on the baseline ($n_{3\sigma}$ > 2). \todo{Specifically, we take the mean, $\mu_b$, and standard deviation, $\sigma_{b}$, of the data points on the baseline (ignoring the reported error bars) and require 3 points $>\mu_{b}+\sigma_{b}$} Secondly, we require \todo{that} the predicted amplification of the PSPL MLE solution at the time of the data point closest to $t_{0}$ to be at least 3 times the standard deviation \todo{($>3\sigma_{b}$)} of the baseline ($\text{Amp}_{\text{sig}}>3$).

\begin{figure}
    \centering
    \includegraphics[width=.9\columnwidth]{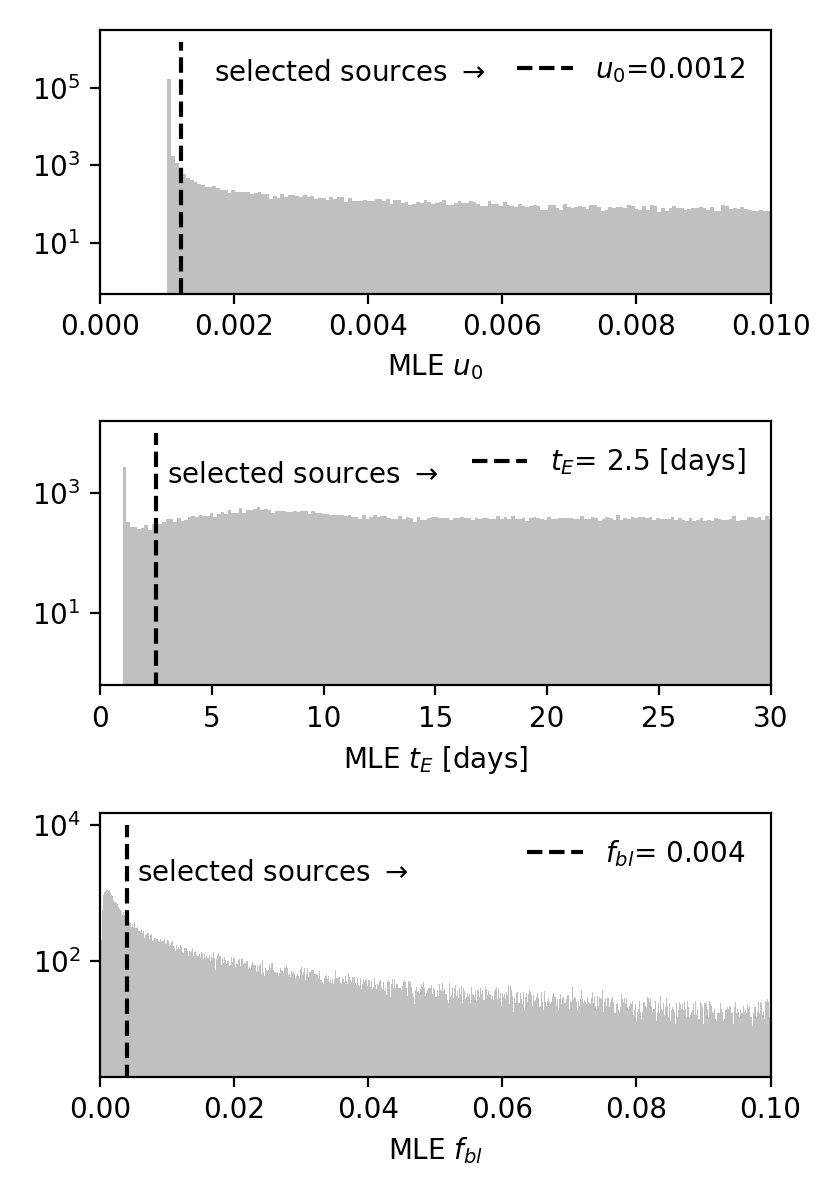}
    \caption{Cuts applied to the maximum likelihood estimate (MLE) parameters for each source. The upper panel is the normalised impact parameter $u_{0}$ and the lower panel is the Einstein timescale $t_{\rm E}$. A build up at the lower optimization boundaries (10$^{-3}$ for $u_{0}$, $1.0$ day for $t_{\rm E}$) are observed. The sources at the boundary are removed with the cuts indicated by the dashed lines. The MLE optimization is highly likely to have failed in these cases.}
    \label{fig:MLE_cuts}
\end{figure}

For the PSPL MLE, optimization bounds for the PSPL parameters had to be chosen (see section \ref{sec:mle}). Naturally, because all the light curves in the survey are being fitted with the PSPL model, most of which are not microlensing, some of the optimizations fail. In some cases, this is indicated by a solution near or on the optimization boundary. Fig.~\ref{fig:MLE_cuts} shows the distribution of the PSPL MLE parameters ($u_{0}$, $t_{\rm E}$, $f_{\rm bl}$) close to the lower end of the optimization boundaries for the 513\,047 sources that met all the other quality cuts described in this section. In each case, a build-up at the lower bound is seen. We remove these sources by requiring the PSPL MLE parameters $u_{0}>0.0012$, $t_{\rm E}>2.5$~days, and $f_{\rm bl}>0.004$. We note that no build-up at the optimization bounds was observed for the $m_{0}$ and $t_{0}$, which were set dynamically for each light curve (see section \ref{sec:mle}), or at the upper bounds for $u_{0}$ or $t_{\rm E}$. An accumulation of sources at the upper bound of $f_{\rm bl}=1$ was also observed. However, we do not eliminate these sources with a quality cut. This is an expected outcome as $f_{\rm bl}=1$ is a reasonable physical solution indicating no blending.

The quality cuts described in this section and the number of sources from the VVV survey that survive them are summarised in Table~\ref{tab:quality_cuts}.  
\subsubsection{Features}
\label{sec:checking_features}

\begin{table*}
	\centering
	\caption{Initial cuts applied to the simulated events and VVV $K_{s}$-band light curve data. The sources remaining are the number of sources surviving the accumulation of the rows above it.}
	\begin{tabular}{l p{120mm} l} 
		\hline
		Cut & Description & sources remaining \\
		\hline
		 None & ... & $669\,244\,443$  \\
		 $n_{\text{before}}> 10$ & number of data points before $t_{0}-2t_{\rm E}$ & $375\,837\,359$\\
		$n_{\text{after}}> 10$ & number of data points after $t_{0}+2t_{\rm E}$ & $159\,192\,879$ \\
		$n_{\text{amp}}> 5$ & number of data points inside $2t_{E,\text{mle}}$ of $t_{0,\text{mle}}$ & $92\,734\,005$ \\
		$n_{3\sigma} > 2$ & number of data points on the amplification above three standard deviations of the baseline & $6\,115\,025$\\
		$\text{Amp}_{\text{sig}}\geq3$ & ratio of the PSPL maximum likelihood solution prediction at the data point time closest to $t_{0}$ and the standard deviation of the baseline. & $513\,047$\\
		$u_{0} > 0.0012$ & maximum likelihood solution for $u_{0}$ is above the lower optimization bound & $241\,648$ \\
		$t_{\rm E} > 2.5$ [days] & maximum likelihood solution for $t_{\rm E}$ is above the lower optimization bound & $228\,985$\\
		$f_{\rm bl} > 0.004$ &  maximum likelihood solution for $f_{\rm bl}$ is above the lower optimization bound & $217\,317$ \\
		\hline
	\end{tabular}
\label{tab:quality_cuts}
\end{table*}

Next we compute model checking features to be used to train the classifier to identify microlensing events. The features are computed from the distribution of standardised residuals between the data and the MLE solution of the PSPL model. Specifically, a standardised residual for a given data point $D_{i}$ is defined as,
\begin{equation}
    \hat{r}_{i} = \frac{m_{i} -m_{\text{PSPL}}(t_{i};\boldsymbol{\theta}_{\text{mle}})}{\sigma_{m_{i}}}.
\end{equation}
Under the assumption that the PSPL MLE solution is the true model and the noise assumptions underpinning the likelihood function in eqn~(\ref{eq:log_like}) are true, the distribution of standardised residuals over all the data should be a Gaussian \todo{with mean zero and unit standard deviation} \citep{Rene2010}. 

We compute the standard deviation of the standardized residuals for data points both inside and outside of two $t_{\rm E}$ of $t_{0}$. This gives two features $\sigma_{\text{amp}}$ and $\sigma_{\text{base}}$, on the amplification and baseline, respectively. The features allow the performance of the PSPL model and its noise modeling assumptions to be evaluated both on and off the microlensing amplification.
\subsection{The von Neumann ratio}

We also compute the von Neumann ratio $\eta$ \citep{von_neumann1941, von_neumann1942}, which is defined as the ratio of the mean square of adjacent differences of data points to the variance of all the data points on the light curve. Specifically, it is
\begin{equation}
\eta = \frac{\sum_{i=1}^{N-1}(m_{i+1}-m_{i})^{2} / (N-1)}{\sum_{i=1}^{N}(m_{i}-\bar{m})^{2}/(N-1)},
\end{equation}
where $N$ is the number of data points on the light curve and $\bar{m}$ is the mean magnitude. Although $\eta$ was derived as an trend indicator under the assumption that the data points are independent draws from a population level Gaussian distribution, it has been demonstrated to be useful in identifying microlensing events \citep[e.g.][]{Pr14, Wy15, Go19}. This is because even if the data aren't distributed according to a Gaussian, $\eta$ still identifies smooth trends as long as the distribution is symmetric or nearly so \citep{Sokolovsky2017, lemeshko2006abbe,strunov2006applying}. In the case of the PSPL model with no noise, the data are not Gaussian distributed, but they are symmetric, suggesting $\eta$ should identify PSPL microlensing events well in high signal to noise regimes. Moreover, $\eta$ is trivially vectorizable and so is cheap to compute for all sources in the VVV survey.

\subsection{Decision Tree Classifier}
\label{sec:decision_tree_classifer}

With our computed features in hand ($\Delta\text{AIC}$, $\sigma_{\text{amp}}$, $\sigma_{\text{base}}$, $\eta$), we train a decision tree classifier to extract microlensing events from the VVV survey. Decision trees are non-parametric supervised machine learning algorithms that classify objects based on computed features and were first proposed by \cite{Breiman1984}. Given a set of objects with computed features and known classes (the training set), a decision tree can be built or trained to predict the class of new data in many ways \citep[see e.g. chapter 9.2 in][]{Hastie2009}. 

\begin{figure}
    \centering
    \includegraphics[width=\columnwidth]{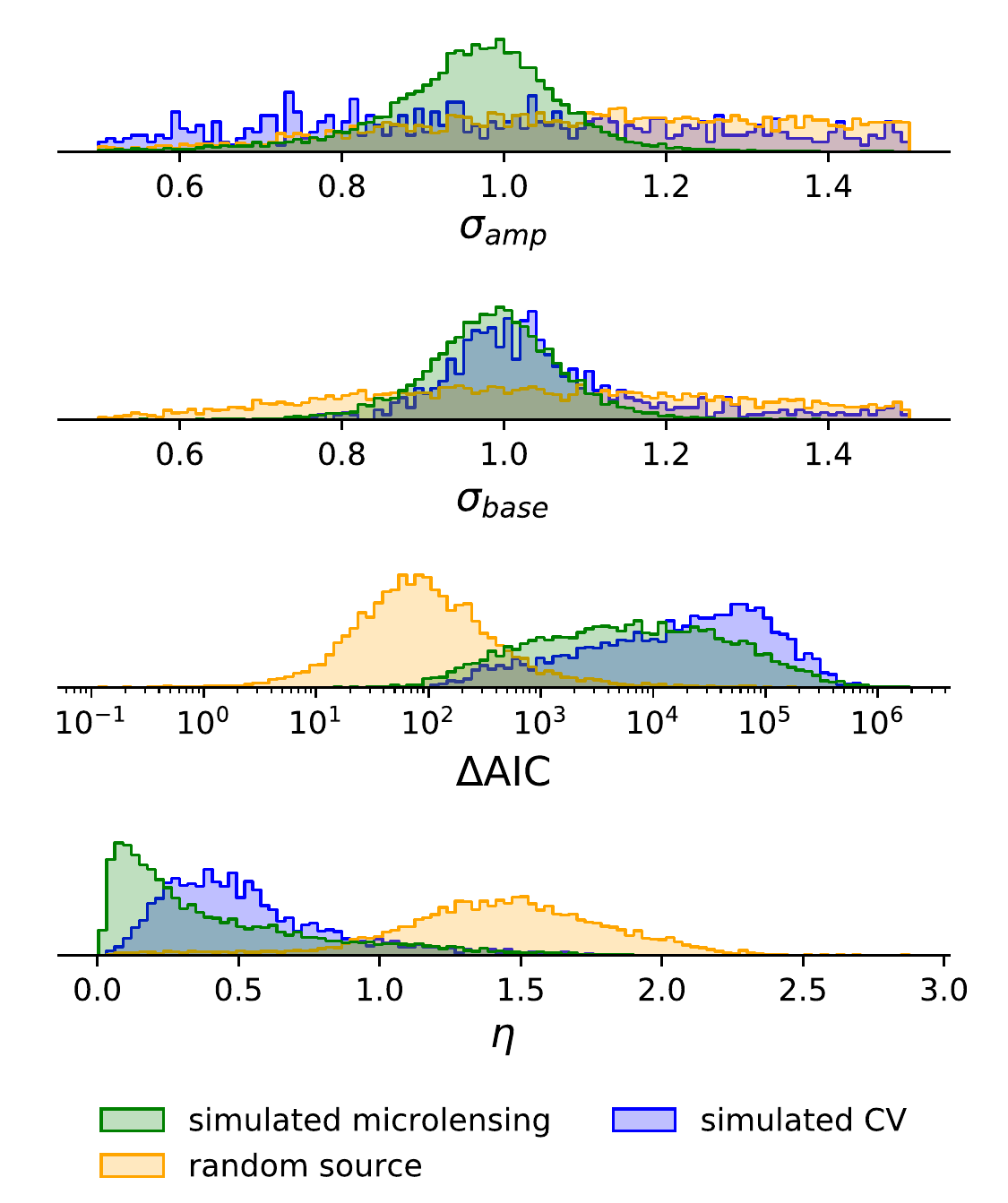}
    \caption{Feature distributions of the training data as input to the decision tree classification model, shown as normalized histograms. The simulated microlensing events are well separated from the random sources in most features. As expected, simulated CVs are less well separated, indicating that they could be a source of confusion for a classifier trained with these features.}
    \label{fig:training_data_features}
\end{figure}

\begin{figure*}
    \centering
    \includegraphics[width=\textwidth]{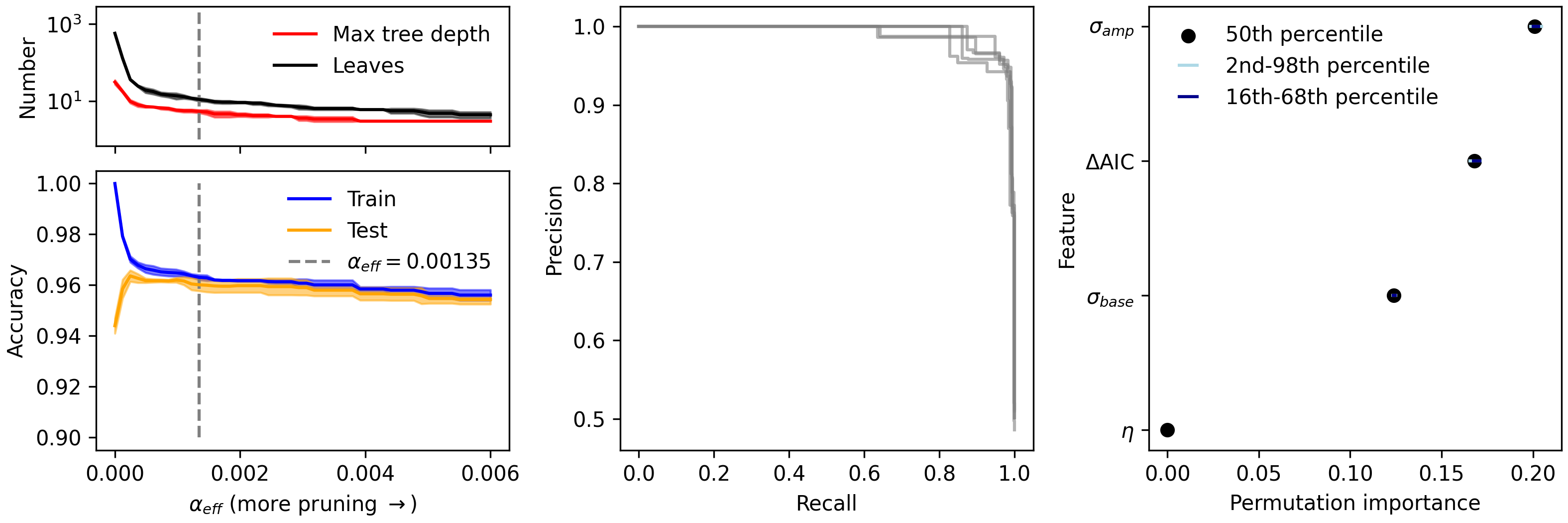}
    \caption{\textbf{Left}: The decision tree accuracy and size as function of how much pruning is performed (increasing $\alpha_{\text{eff}}$). In both panels, coloured bands show the mean and standard deviation of each metric over a 5-fold cross validation. \todo{$\alpha_{\text{eff}}=0.00135$} (dashed grey line) is the smallest value chosen such that tree performance on the training data (blue) is at least one standard deviation above the tree performance on the test data (orange). If there is less pruning than this threshold, the tree starts to over-fit on the training data. \textbf{Middle}: Precision recall curve of the pruned decision tree. Grey lines show the curve for each fold of a 5-fold cross-validation. \textbf{Right}: Permutation importance of each of the features. This shows how much the tree relies on each feature to classify microlensing events. The $\sigma_{\text{amp}}$ feature is the most important discriminator followed by $\Delta\text{AIC}$.}
    \label{fig:decision_tree_tuning}
\end{figure*}

We use a training algorithm similar to C4.4 \citep{quinlan2014}, which is implemented in the Scikit-learn Python package \citep[namely \textsc{DecisionTreeClassifier};][]{scikit-learn}. Let $Q_{i}$ represent the set of features and known class label (microlensing or not microlensing) for one example in the training data. Then the set of all $N_{T}$ training examples is $\mathcal{Q} =\{Q_{i}\}_{i=1}^{N_{T}}$. In this algorithm, at the root node in the tree,§ there is a binary split in one feature which best separates the training data into their target classes in the two generated left and right child nodes. This split is chosen to minimize the total Gini impurity measure $I$ across both child nodes. Specifically, for a given split in the chosen feature, we have
\begin{equation}
    I =\frac{N_{\text{left}}}{N}G\left(\mathcal{Q}_{\text{left}}\right) + \frac{N_{\text{right}}}{N}G(\mathcal{Q}_{\text{right}})
\end{equation}
Here, $\mathcal{Q}_{\text{left}}$ and $\mathcal{Q}_{\text{right}}$ are the subsets of $\mathcal{Q}$ that end up in the left and right child nodes after the split, containing $N_{\text{left}}$ and $N_{\text{right}}$ training examples, respectively. Initially, $N_{T}=N_{\text{left}}+N_{\text{right}}$ is the total number of training samples being considered at the parent node. $G$ is the Gini impurity measure computed for the data in each child node,
\begin{equation}
    G(\mathcal{Q}_{s}) = p_{\text{m}} (1 - p_{\text{m}}) + p_{\text{nm}} (1 - p_{\text{nm}}).
    \label{eq:gini}
\end{equation}
Here, $p_{\rm m}$ and $p_{\rm nm}$ are the proportion of light curves that are labeled as microlensing and not microlensing respectively, in subset $Q_{s}$. This procedure is recursively repeated for each new child node containing a subset of the data and the tree grows in size, until some stopping criteria is reached. The stopping criteria can be a maximum tree depth, or a minimum number of training samples or a minimum threshold impurity decrease in each of the proposed child nodes.

Once the stopping criteria has been reached, all nodes in the tree without children are called leaf nodes. The trained decision tree can then be used to predict the class of an object from the leaf node the object ends up in. For each leaf node, the probability that the object is a particular class is the proportion of that class that ended up in that leaf node during training. In our case, for each leaf node in the tree, $p_{\rm m}$ is the probability that the object is microlensing and $p_{\rm nm}$ not microlensing. For an object to be initially classified as microlensing, its microlensing class probability has to be $> P_{\text{thresh}} = 0.5$. It is worth noting that since the training set does not contain realistic proportions of simulated events and contaminants, in addition to a pre-selection with quality cuts, the probabilities extracted are not a true measure of the probability of a particular class. However, the values are still useful because we expect leaf nodes with higher probabilities to contain more genuine microlensing events than leaf nodes with lower probabilities.

The advantages of using a decision tree classification algorithm are that it does not require pre-processing of the input features; once trained it is fast and scalable to run (prediction has logarithmic cost in the number of data points used to train the tree); and it is easy to interpret (prediction can be explained by Boolean logic) when compared with other classification algorithms such as neural networks. Disadvantages are that it tends to over-fit training data, unless regularized or pruned in someway; it is also poor at interpolating and does not form smooth decision boundaries.\footnote{See the scikit-learn documentation on decision trees (\url{https://scikit-learn.org/stable/modules/tree.html}) and references therein for a detailed explanation of the advantages and drawbacks of decision trees.}

For our training data, we used 9\,000 simulated microlensing events and 2\,250 simulated CVs from the simulation procedure detailed in section \ref{sec:simulations} that pass all the quality cuts in Table \ref{tab:quality_cuts}. We also take 6\,750 random light curves from the VVV data (which are vanishingly unlikely to be true microlensing events) which also pass all the quality cuts. We label both the CV simulations and random selection of light curves to be members of the not microlensing class. Fig.~\ref{fig:training_data_features} shows the distribution of the training data in the features with the classes overlaid. The microlensing simulations are well separated from the random light curves in most of the features. However, the CVs are less well separated, indicating they will likely be a source of contamination. 

There are many possible strategies to train and tune the decision tree classifier. Typically, stopping criteria (e.g. maximum tree depth or minimum number of samples required to be in a leaf node) are tuned as hyper-parameters of the classifier and are chosen to maximise the classification accuracy. This can be done by an expensive grid search over all of the hyper-parameters \citep[e.g.][]{Go19} to find an optimal combination. We found that this method tended to choose large trees, (depths of 20-30 levels) with many parts that only slightly increase classification performance. 

In order to keep the decision tree small and scalable and to prevent over-fitting on the training data whilst keeping its accuracy high, we adopt a two-stage training procedure. We first train the decision tree with fixed hyper-parameters that do not limit the size of the tree. Specifically, we do not limit the maximum tree depth, we don't require any threshold minimum impurity decrease at a split which only needs two training examples, and we only require one training example to be present in each leaf node. We then prune the tree and remove nodes that don't significantly affect the classification accuracy using minimal cost complexity pruning \citep{Breiman1984}.

Minimal cost complexity pruning aims to find the subtree of the unpruned tree $T$ that minimizes the cost complexity measure 
\begin{equation}
    R_{\alpha}(T) = R(T) + \alpha|\Tilde{T}|.
\end{equation}
Here $R(T)$ is the total sample weighted impurity of all the leaf nodes in $T$, $\alpha\geq0$ is the cost complexity parameter, and $|\Tilde{T}|$ is the total number of leaf nodes in $T$. The cost complexity of a single node $t$ in the tree is similarly defined as $R_{\alpha}(t) = R(t) + \alpha$, where $R(t)$ is the impurity measure at the node. For a given branch of the tree $T_{t}$ with root node $t$ the effective cost complexity $\alpha_{\text{eff}}$ is defined by setting $R_{\alpha}(T_{t})=R_{\alpha}(t)$,
\begin{equation}
    \alpha_{\text{eff}} = \frac{R(t)-R(T_{t})}{|\Tilde{T_{t}}|-1}.
\end{equation}
For each node in the tree, excluding leaf nodes, this quantifies how well the subtree below the node decreases the classification impurity, regularised for the size of the subtree. Nodes with a high $\alpha_{\text{eff}}$ are the top of subtrees which decrease the classification impurity efficiently for their subtree size and should be kept. Pruning is performed by choosing a threshold value for $\alpha_{\text{eff}}$ for which all nodes in the tree that have $\alpha_{\text{eff}}$ value lower than the threshold are removed. Overall, pruning decreases the size of the tree by removing the weakest parts which do not contribute significantly to discriminating between classes for their size.

Fig.~\ref{fig:decision_tree_tuning} (left panels) shows the decision tree average classification accuracy over the microlensing and non-microlensing classes as a function of the threshold value of $\alpha_{\text{eff}}$ (the amount of pruning). Accuracy is calculated over a 5-fold cross validation on the 18\,000 training samples. For each fold, the accuracy on the held out data (test) and on the training data (train) is evaluated. A value of training accuracy that is higher than test accuracy suggests that the decision tree is over-fitting. We therefore choose the value of threshold \todo{$\alpha_{\text{eff}}=0.00135$} which corresponds to the point where the train accuracy become significantly higher than the test accuracy. Specifically, over the 5-fold cross validation, this is the point where the mean train accuracy is greater than one standard deviation above the mean test accuracy. This pruning results in a significantly smaller and simpler tree (< 10 levels deep compared with the unpruned 20-30 levels), that is not over-fitting, and retains a high accuracy$~\sim96$ per cent on the test data. The threshold \todo{$\alpha_{\text{eff}}=0.00135$} is fixed and the tree is trained on all 18\,000 training samples to produced the final trained classifier. 

Fig.~\ref{fig:decision_tree_tuning} (middle panel) shows the precision recall curve for the decision tree classifier. Precision and recall are defined as,
\begin{equation}
    \text{Precision} = \frac{\text{true positives}}{\text{true positives} + \text{false positives}}, \label{eq:precision}
\end{equation}
\begin{equation}
    \text{Recall} = \frac{\text{true negatives}}{\text{true negatives} + \text{false positives}}. \label{eq:recall}
\end{equation}
Here, a true positive is the correct classification of a microlensing event. The middle panel of Fig.~\ref{fig:decision_tree_tuning} shows how these quantities vary as the $P_{\text{thresh}}$ is increased above 0.5. Requiring higher $P_{\text{thresh}}$ trades off a purer sample of microlensing events (high precision) against a more complete sample (high recall). The precision recall curve in Fig.~\ref{fig:decision_tree_tuning} suggest both a high level of completeness and purity can be achieved with the classifier on the training data by choosing a $P_{\text{thresh}}$ corresponding to a point in the top right corner of the plot. We will however return to this point when we apply the classifier over the real dataset in Section~\ref{sec:results}.

The right panel of Fig.~\ref{fig:decision_tree_tuning} shows the permutation importance for each feature. This is the change in classification accuracy of the decision tree when the feature is corrupted by random permutation \citep{feat-importance}. This is repeated a set number of times (10 in this work) for each feature, giving rise to a distribution of permutation scores. A feature with a high permutation importance is relied on more by the decision tree to produce accurate classifications. Fig.~\ref{fig:decision_tree_tuning} shows that $\sigma_{\text{amp}}$ is most relied on by the decision tree. We note that $\eta$ has a permutation importance of zero, but this does not mean that the decision tree does not use it at all. The inclusion of $\eta$ does not increase the average classification accuracy when considering purely a crossing of the threshold $P_{\text{thresh}}=0.5$, but we found it did increase the purity of the \todo{$p_{m}$}$>0.5$ bins.

\subsection{Validation}

To evaluate the performance of the trained classifier, it is applied to a previously unseen validation dataset distinct from the training data used to train and tune the classifier in Section~\ref{sec:decision_tree_classifer}. This data set contains 2\,000 examples in the same proportions and labelling as the training data (1\,000 simulated microlensing events, 333 simulated CVs, and 667 random light curves -- or 1\,000 non-microlensing events).

\begin{figure*}
    \centering
    \includegraphics[width=\textwidth]{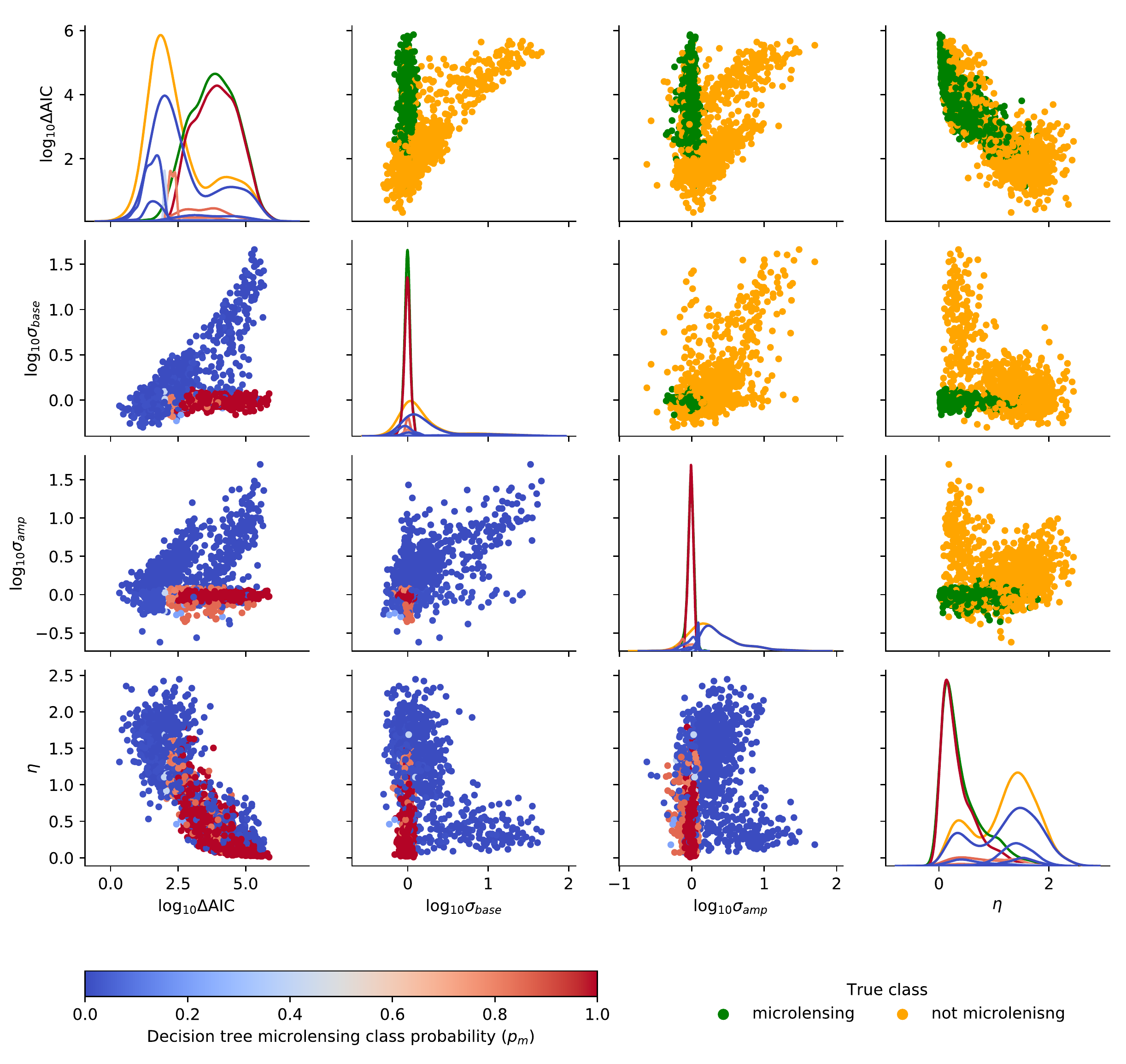}
    \caption{Projected decision surface for the trained decision tree evaluated on the validation data set. Lower scatter plots show the validation data coloured by the \todo{microlensing} class probability they are assigned by the trained decision tree \todo{($p_{m}$)} in all two dimensional projections of the feature space. Upper scatter plots show the same two dimensional projections in the feature space but the sources are coloured by their true classes.}
    \label{fig:decision_surface}
\end{figure*}

Fig.~\ref{fig:decision_surface} shows two-dimensional projections of the decision boundary, with the lower left panels coloured according to the verdict of the classifier and the upper right panels coloured according to the known classification. The diagonal panels show the one-dimensional distributions of these classes for each feature. The fact that the red and green lines and data points -- actual microlensing and claimed microlensing, respectively -- shadow each other closely in both the one and two-dimensional distributions is very encouraging. The decision boundary in feature space is clearly complex, but there are prominent regions even in the projected spaces in which the microlensing events clearly predominate.

The performance of the classifier on the validation set can be made quantitative by computing the confusion matrix, as in \todo{Table~\ref{tab:confusion_matrix}}. The classifier achieves 97 per cent accuracy on identifying microlensing events, and 94 per cent on identifying non-microlensing. This is very respectable, however the rate of $\sim{}6\%$ of non-microlensing samples misclassified as microlensing may yet prove troublesome.
If the actual data had the same proportions of microlensing events to cataclysmic variables (the principal contaminants) as the validation set, then this would still yield substantial numbers of false positives on large datasets. 

\todo{The frequency of CVs can be estimated from the Gaia Catalogue of Nearby Stars~\citep{Ga21}, which contains $\sim 320,000$ stars. The local space density of CVs is $\sim 2 \times 10^{-5}$ \citep{Ak08} which suggests that their frequency is $\sim 6 \times 10^{-5}$, or roughly an order of magnitude greater than microlensing. This though is an overestimate as to the seriousness of the contamination problem in VVV data because (i) the definition of CVs includes many classes of dwarf novae (such as SS Cygni stars) which repeat on timescales of a few weeks and so would not be mistaken for microlensing, and (ii) the amplitude of the outburst decreases with increasing wavelength and so the problem is less severe in the infrared VVV as opposed to optical surveys like OGLE. The troublesome CVs have rather specific characteristics -- the CV must rise unusually slowly or the data points corresponding to the rise must be missing. Such troublesome CVs are very much less frequent than microlensing.}

\begin{table}
    \centering
    \begin{tabular}{l||ll}
         &  \multicolumn{2}{c}{Predicted class}  \\
    True class   & Microlensing & Not microlensing \\
        \hline \\
    Microlensing &  $97\%$ & $3\%$ \\
    Not microlensing & $6\%$  & $94\%$ \\
    \hline
    \end{tabular}
    \caption{\todo{Confusion matrix on the previously unseen held out validation data of $2000$ examples ($1000$ microlensing, $1000$ not microlensing). Values in the confusion matrix are the fraction of validation examples normalised by row.}}
    \label{tab:confusion_matrix}
\end{table}


\section{Results and Completeness analysis}
\label{sec:res}

\subsection{Results} 
\label{sec:results}

Once the classifier has been trained, it is applied to the 217\,317 VVV light curves that passed the quality cuts in Table~\ref{tab:quality_cuts}.
This yields $6\,677$ microlensing candidates with a decision tree microlensing class probability $> 0.5$. Many of these are not microlensing, but this is not surprising.  The training set consists only of the order of $10^4$ examples, so there are bound to be cases of contamination and noise on which the classifier has not been trained. Additionally, a rate of $\sim{}6\%$ of non-microlensing training samples misclassified as microlensing (see the confusion matrix in \todo{Table \ref{tab:confusion_matrix}}) translates to $\lesssim 10^4$ false positives in our light curve set -- though the frequency of the major contaminant (cataclysmic variables) is likely over-represented in the training and validation sets as compared to the actual VVV data, so this calculation is unduly pessimistic. 
 
\begin{figure}
    \centering
    \includegraphics[width=\columnwidth]{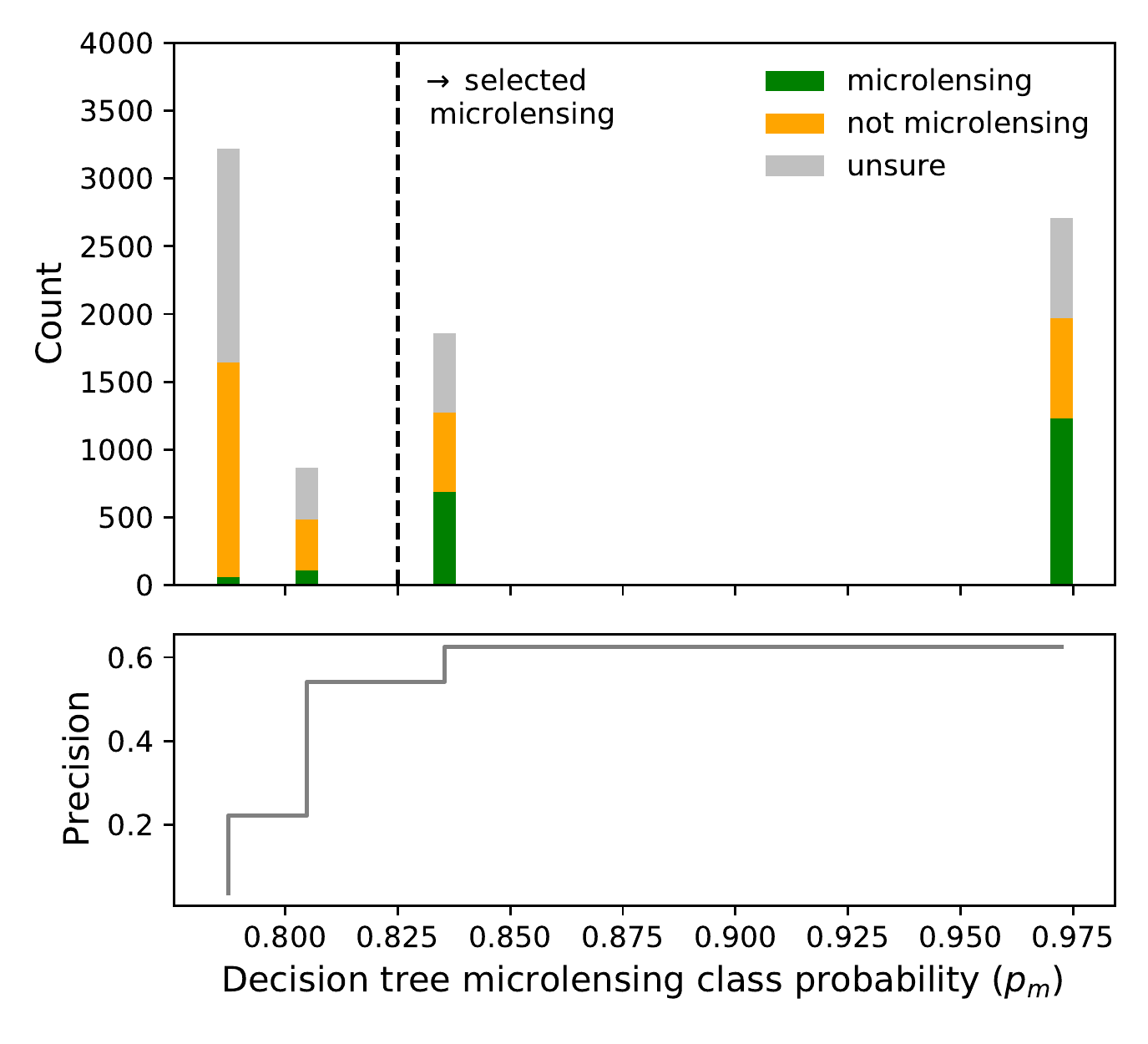}
    \caption{Actual purity of the sources with microlensing class probability \todo{$p_{m}$}$\geq0.5$. which are classified by visual inspection of the light curves. Dashed line indicated the threshold probability ($P_{\text{thresh}} = 0.825$) used to select microlensing events for the final sample. Values of precision in this plot are calculated ignoring sources with an unsure visual classification.}
    \label{fig:actual_purity}
\end{figure}

Nonetheless, a visual inspection was then required to remove the remaining false positives. First, a preliminary classification of all $6\,677$ light curves was made by one of the authors to produce Fig.~\ref{fig:actual_purity}. The top panel shows the ratio of microlensing to non-microlensing as a function of the decision tree microlensing class probability \todo{$p_{m}$}, whilst the bottom panel shows the purity, as determined from this first visual inspection. Using the flattening of the purity curve as a guide, we chose $P_{\text{thresh}}\todo{=}0.825$ to produce a set of $4132$ light curves for further scrutiny with the aim of refining to produce a very pure sample. 

Each of the $4\,132$ microlensing candidates was now visually inspected by at minimum a further two authors (meaning at least three of the authors saw each candidate), who adopt the scoring of $-1$ for not microlensing, $0$ for unsure and $1$ for microlensing. We decided that light curves with a mean score $\geq0.5$ were almost certainly genuine microlensing events, yielding a final list of $1\,959$. This catalogue is presented in electronic form in Appendix A as Table~\ref{tab:event_inference}. Some sample light curves that are classified as microlensing are displayed in Fig.~\ref{fig:mosaic}, together with results from our later maximum likelihood and Bayesian modelling (see Section~\ref{sec:nested_sampling}). \todo{In Sections \ref{sec:comp} and \ref{sec:tE_inference} we analyse the final sample of $1\,959$ human-vetted events.}

\subsection{Visual inspection} \label{sec:vis_inspection_test_outcome}

Spread randomly among the 4\,132 real light curves were 150 simulated events that had passed the classifier. We included these to gauge completeness, with each participant seeing the same set of simulations. In addition, since the visual inspection was performed in two rounds, 100 event candidates that each participant had seen in the first round were also included in the second round to test consistency.

The completeness of the visual inspection as determined by the recovery of the simulated microlensing was particularly surprising. The most complete participant still missed $23$ per cent of the simulated events, the least complete missed $70$ per cent of them, the remaining two missed $27$ and $51$ per cent. The large variation in completeness between participants was astonishing to us. Completeness was most highly correlated with baseline magnitude and $u_0$, with higher values yielding lower completeness. This is not surprising since bright, high amplitude events have the clearest signals. Considering simulated events having a mean score $\ge{}0.5$ as above, then our visual inspection recovers 73 of 150 events, or an overall completeness of just $50$ per cent!

The consistency of three of the participants was such that only $\sim{}6$ per cent of classifications were significantly different between rounds, where \lq  significantly different' is deemed to be a switch from microlensing to non-microlensing, or vice-versa. One participant was inconsistent at the $20$ per cent level. We note that the direction of change in disposition of the participants also varied. Two participants clearly became more strict in the second round of inspections (i.e. where a change in classification occurred it was generally towards non-microlensing), one was only marginally more strict, and the last was more relaxed relative to the first round.

In conclusion, visual inspection is likely to yield time-varying results with poor completeness. We partially mitigated the effects by having 4 inspectors. For future surveys, we advocate having far more inspectors view each candidate and taking an aggregate score. This could be achieved with a citizen science project which allows tens of thousands of volunteers to classify astrophysical objects \citep[e.g.,][]{Lintott} or photometric time series data \citep[e.g.][]{Thiemann2021}. This should provide robust and reliable classifications. Unfortunately, the results of the completeness test indicate that visual classification is no guarantee of a high level of completeness. Even with a large number of inspectors, there is likely to be a point beyond which higher completeness cannot be reached. It is better to design a classification method which requires no human intervention at all to produce a final (incomplete) catalogue of events, though even this is difficult to do for such an intrinsically rare phenomenon as microlensing. However, it is then much easier to quantify completeness and contamination of an exclusively machine-classification.

\subsection{Spatial variation}\label{sec:spatial_completeness}

\begin{figure*}
    \centering
    \includegraphics[width=\textwidth]{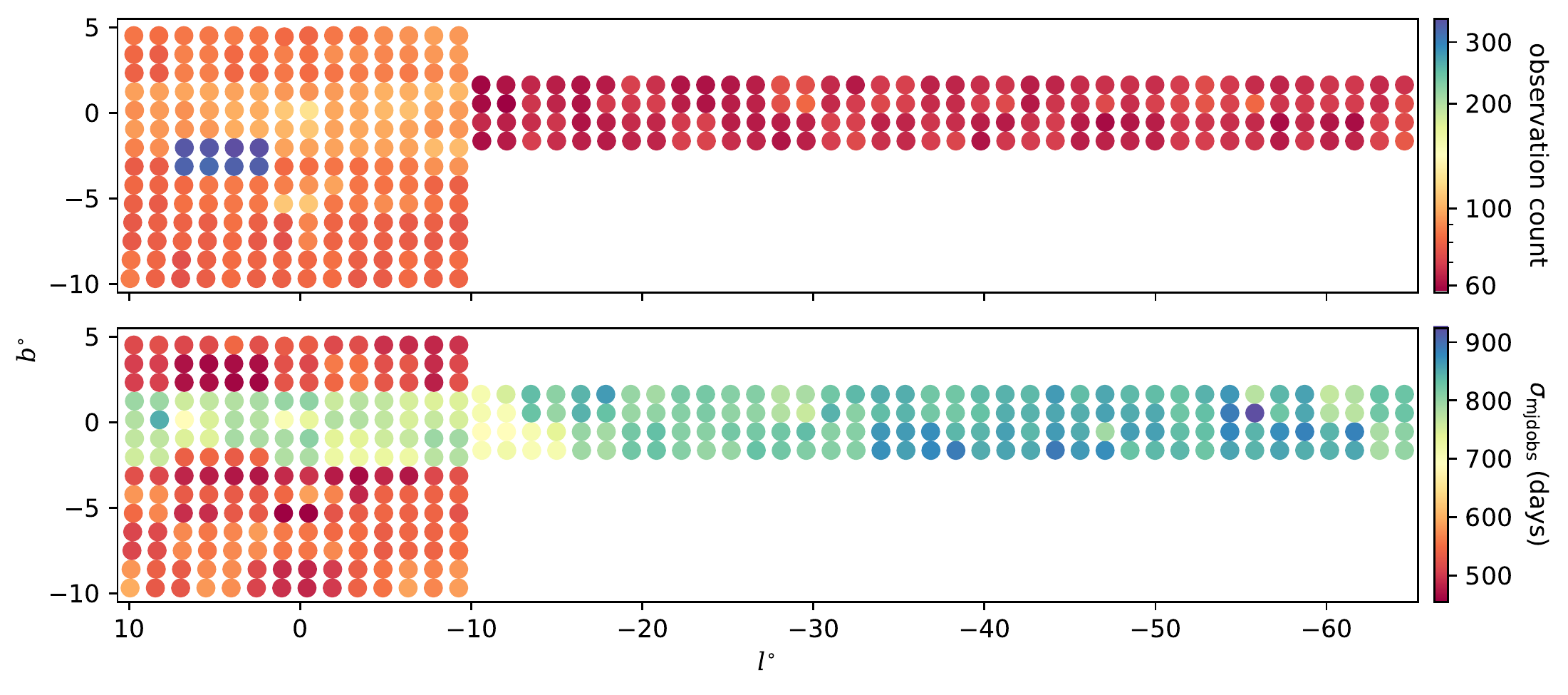}
    \caption{The spatial and temporal density and distribution of the VVV survey, including VVVX observations that cover the original VVV survey area. The upper panel colour axis shows the number of observations per pointing position for a given tile (note the logarithmic scaling for this panel), and the lower panel shows the standard deviation in observation date in Julian days.}
    \label{fig:obs_distribution}
\end{figure*}

Spatial variation in transient detection efficiency is to be expected when observations are sparse and irregularly sampled. The VVV survey observation strategy is split between four different zones (spatial ranges approximated):
The Galactic disc fields at $|b|<2^{\circ}$ and $295<l^{\circ}<350$ have $\sim{}65$ observations per pointing spread relatively evenly across the 10 year survey;
The mid-plane bulge fields at $|l-0^{\circ}|<10^{\circ}$ and $|b|<2.5^{\circ}$ have $\sim{}95$ observations per pointing, with the majority in the calendar years 2012 to 2015 (inclusive);
The high latitude Galactic bulge fields at $|l-0^{\circ}|<10^{\circ}$ and $|b|>2.5^{\circ}$ have $\sim{}80$ observations per pointing with a similar spread to the disc fields; 
The high cadence region in the inner bulge at $1.5<l^{\circ}<7.5$ and $-3.5<b^{\circ}<-1.5$ has $\sim{}335$ observations per pointing, with the majority being in the 2012, 2014 and 2015 calendar years.
It should be noted that these counts are per pointing, but the VIRCAM tiling strategy means sources are covered by between 1 and 6 pointings (2 being the modal value). This means a randomly selected position within the survey may be observed anywhere between $\sim{}65$ and $\sim{}2000$ times. The observation count of positions separated by an arcminute can differ by an order of magnitude. The spatial distribution of VVV tiles, coloured by both a per-pointing observation count and the standard deviation of the date of their observation (in days) are shown in Fig.~\ref{fig:obs_distribution}.

\begin{figure*}
    \centering
    \includegraphics[width=\textwidth]{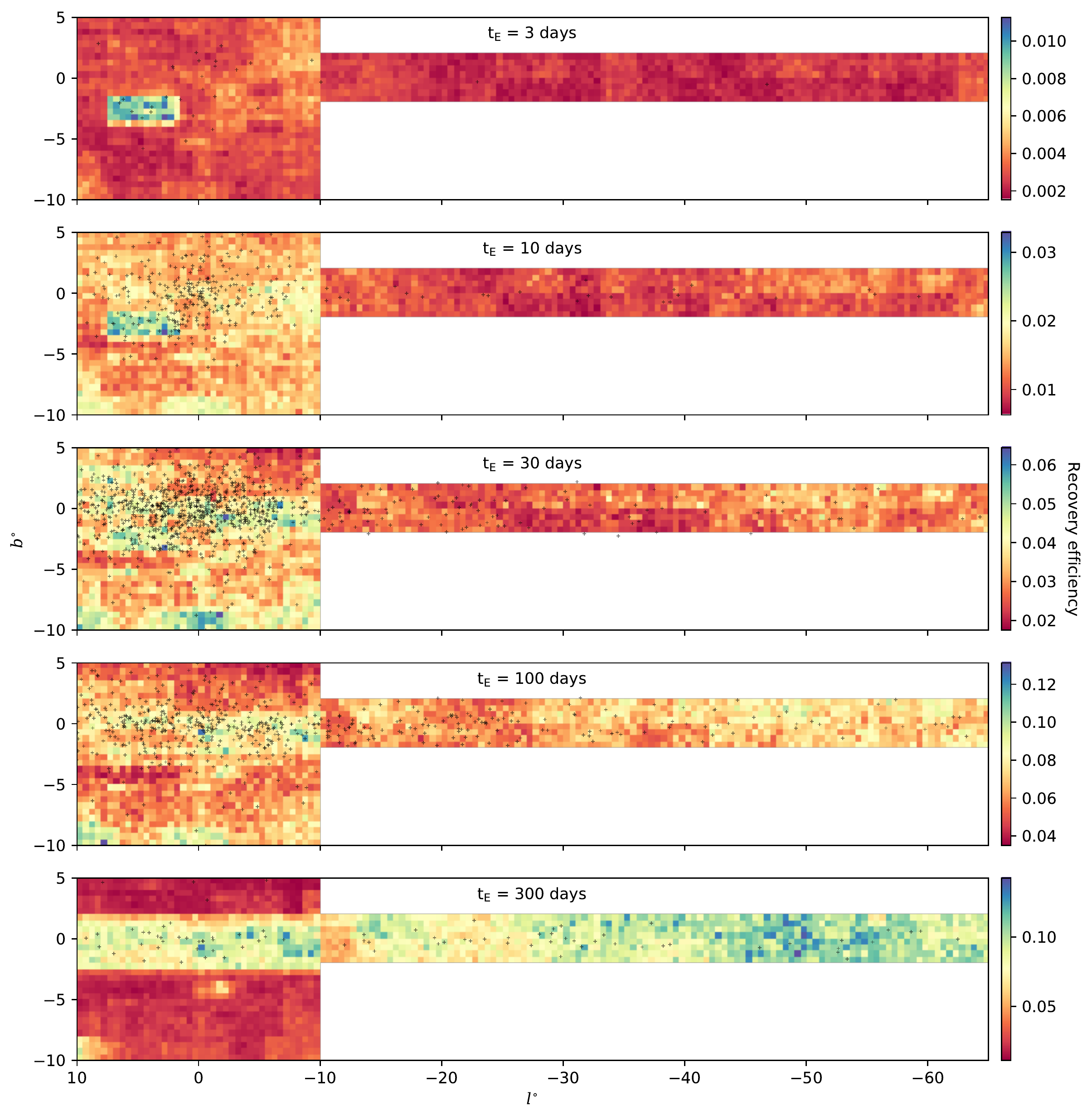}
    \caption{Simulated microlensing event recovery efficiency for 5 different values of Einstein time and for $10^{-3}<u_0<1.5$. For each spatial bin events are simulated and classified, and the recovery efficiency is the fraction of events classified as microlensing. There is a $10\%$ uncertainty on these recovery efficiency values. The black points indicate the position of recovered microlensing events, their median $t_{\rm E}$ dictates the map on which they are placed. These maps highlight the dependence of event recovery efficiency on observation cadence and distribution. The high cadence VVV fields in the bulge at $l,b\approx{}5,-2^{\circ{}}$ are clearly visible with a higher recovery efficiency for small $t_{\rm E}$ but are less powerful for longer timescale events. The band of high recovery efficiency at $|b|<2^{\circ}$ in the $t_{\rm E}=300$~day map corresponds to a broader spread in observation dates for these fields.}
    \label{fig:recovery_efficiency}
\end{figure*}

To evaluate the impact of the spatial variation in observation distribution on the microlensing event recovery efficiency we measured the fraction of simulated microlensing light curves recovered by our \todo{classifier (events with $p_{m}>0.825$)} as a function of sky position and Einstein crossing time. The VVV survey area was binned into $0.5^{\circ}\times{}0.5^{\circ}$ sections, with simulation seed sources selected randomly from within each section. Microlensing event parameters used a discrete value of $t_{\rm E}$, a $u_0$ value randomly selected from $\mathcal{U}(10^{-3}, 1.5)$ and the remaining parameter distributions as in Sec.~\ref{pspl_model}. Simulated events were run through our classifier until the recovery efficiency -- the fraction of events classified as microlensing relative to the simulation count -- reached a signal-to-noise ratio of $10$. This experiment was repeated for a total of $5$ values of $t_{\rm E}$. The resulting recovery efficiency maps are shown in Fig.~\ref{fig:recovery_efficiency} with the positions of real recovered microlensing events included in the map with the closest $t_{\rm E}$ value.

We note that the regions covered by $N\neq{}2$ pointings are not clearly visible in Fig.~\ref{fig:recovery_efficiency} as they are much smaller in size than our spatial bins. Observations from the same VIRCAM tile visit are generally all within a few minutes of each other, hence they do not significantly alter the epoch distribution. Consequently, their influence on the light curves for events of a much longer timescale than this is simply to more tightly constrain the magnitude at that time point, typically by a factor $\sqrt{N}$. Their contribution to the completeness efficiency of each is still included in these maps, as the value for each bin is an average efficiency across the whole bin.

\begin{figure*}
    \centering
    \includegraphics[width=\textwidth]{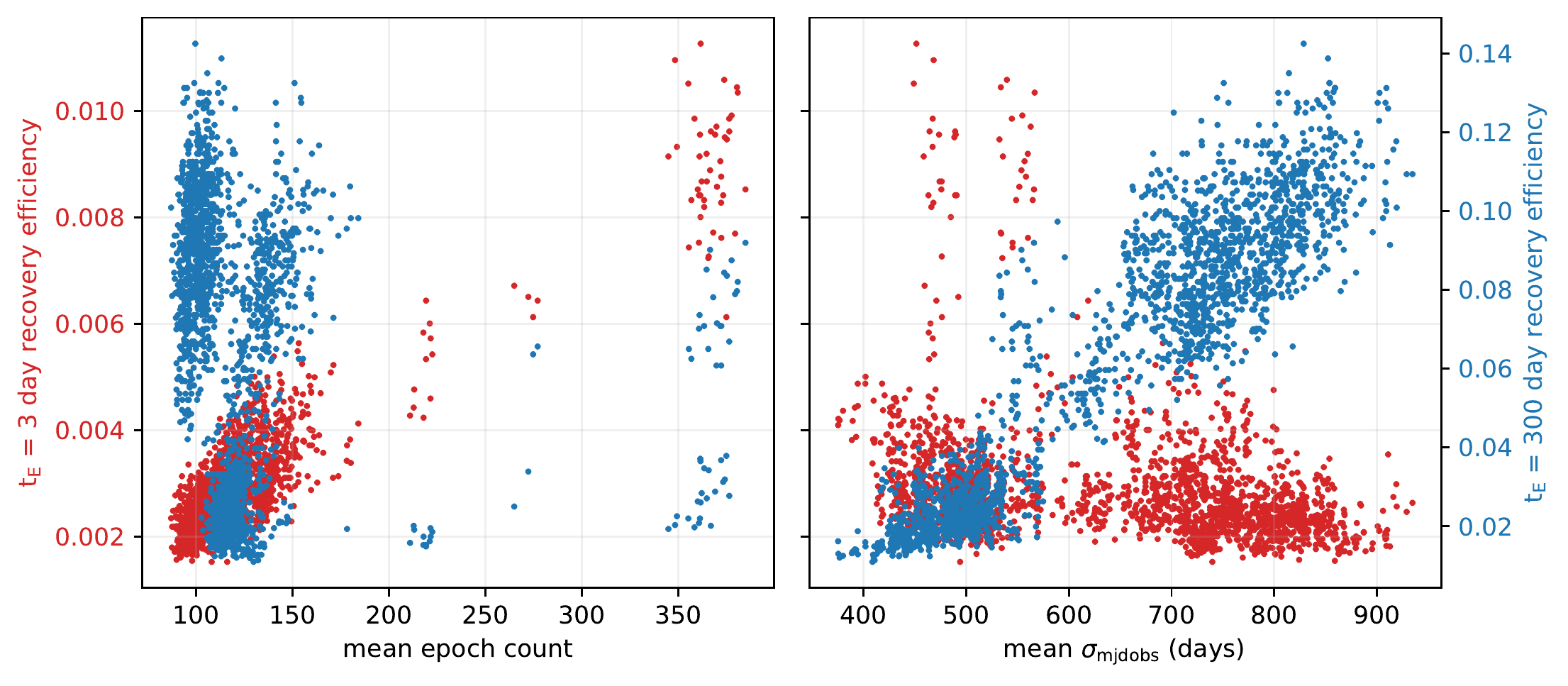}
    \caption{Simulated microlensing event recovery efficiency for ${\rm t}_{\rm E}=3$~day and ${\rm t}_{\rm E}=300$~day events as a function of mean epoch count and mean observation temporal spread of sources in each spatial bin as described in Section~\ref{sec:spatial_completeness}.}
    \label{fig:rec_eff_obs_stats}
\end{figure*}

A cursory inspection of Figures \ref{fig:obs_distribution} and \ref{fig:recovery_efficiency} indicates that high recovery efficiency for low $t_{\rm E}$ events depends more on observation density than it does for high $t_{\rm E}$ events, for which large observation spread is more important. The high cadence region, which has high observation density but small spread, is very evident as a region of (relative) high efficiency in the $t_{\rm E}\le{}10$~day panels but largely blends into the background in the high $t_{\rm E}$ plots. Conversely the entire mid-plane ($|b|\lesssim{}2.5^{\circ}$) has higher efficiency for long timescale events than the high latitude fields due to its larger relative temporal spread. A number of smaller features are evident in the various recovery efficiency maps in Fig.~\ref{fig:recovery_efficiency}, all of which appear to be mirrored to some degree in at least one of the maps of observation density and spread shown in Fig.~\ref{fig:obs_distribution}. The relationship between event recovery efficiency and observation density and distribution for long and short timescale events is clearly visible in Fig.~\ref{fig:rec_eff_obs_stats}, which shows the mean epoch count and mean standard deviation in observation date (in days) for all sources in each spatial bin shown.

The absolute values of recovery efficiency vary hugely, ranging from $<0.2$ per cent for $t_{\rm E}=3$~days in the fields with the lowest observation counts, to $>14$ per cent for $t_{\rm E}=300$~days in the fields with the highest temporal spread in their observations. Even within single values of $t_{\rm E}$ the efficiency can vary by up to an order of magnitude depending on position. This reinforces our view that there is a need to take great care when estimating microlensing survey completeness where inferences on the underlying event rate are to be made.

\section{Comparison with previous VVV microlensing searches}
\label{sec:comp}

Here, we compare our recovered sample of microlensing events to the three catalogues of previously found microlensing events in the VVV \citep[][hereafter the Navarro sample]{NavarroLatitude, NavarroForesaken, NavarroFardisk}. First, we report on the number of previously known microlensing events our decision tree classifier recovers. We then introduce our Bayesian methodology for inferring the PSPL parameters for each microlensing event. We compare our Bayesian approach with the previously used fitting methods for inference of PSPL microlensing parameters in the VVV.

\subsection{Known event recovery}

The Navarro sample reports $855$ unique microlensing events, of which 526 survive the quality cuts outlined in Section~\ref{sec:quality}. Finally, $256$ of those make it into the catalogue presented in this paper. In the end, our catalogue contains $30$ \todo{per cent} of the reported microlensing events in the Navarro sample.

This relatively low recovery fraction highlights the trade-off between careful visual inspection of all light curves over a small area of the survey versus a more automated scalable search over the full survey area. Both methods have advantages and drawbacks. In the case of a visual inspection, almost all the high amplitude microlensing events will be found. For the low amplitude, noisy events, our testing and experimentation detailed in Section~\ref{sec:res} suggests that performance can be patchy and subjective. A careful visual search could provide a somewhat complete sample, at the cost of lack of purity, but this method is not scalable to the $\sim10^{9}$ light curves in the survey. This limits analysis to relatively small numbers of events. Moreover, relying entirely on visual inspection introduces biases into the selection of events that are hard to quantify. This causes difficulties when subsequent population analyses of events is performed -- for example, optical depth or timescale distribution calculations.

In contrast, an automated search scales to all $\sim10^{9}$ light curves in the survey, allowing many more events uniformly over the survey to be found. As demonstrated by the low recovery fraction, this is at the cost of completeness. The real strength of the method is the ability to \todo{better} quantify the selection \todo{function} as shown in Section~\ref{sec:spatial_completeness}, which is vital for subsequent event population analyses. Though the catalogue presented in this work is likely less complete than the Navarro sample, it has the advantage of a \todo{comparatively better quantified} selection function. \todo{This is because we have minimized as much as possible the human component of selecting events, and precisely characterised the selection function of the decision tree classifier.}  

\begin{figure}
    \centering
    \includegraphics[width=\columnwidth]{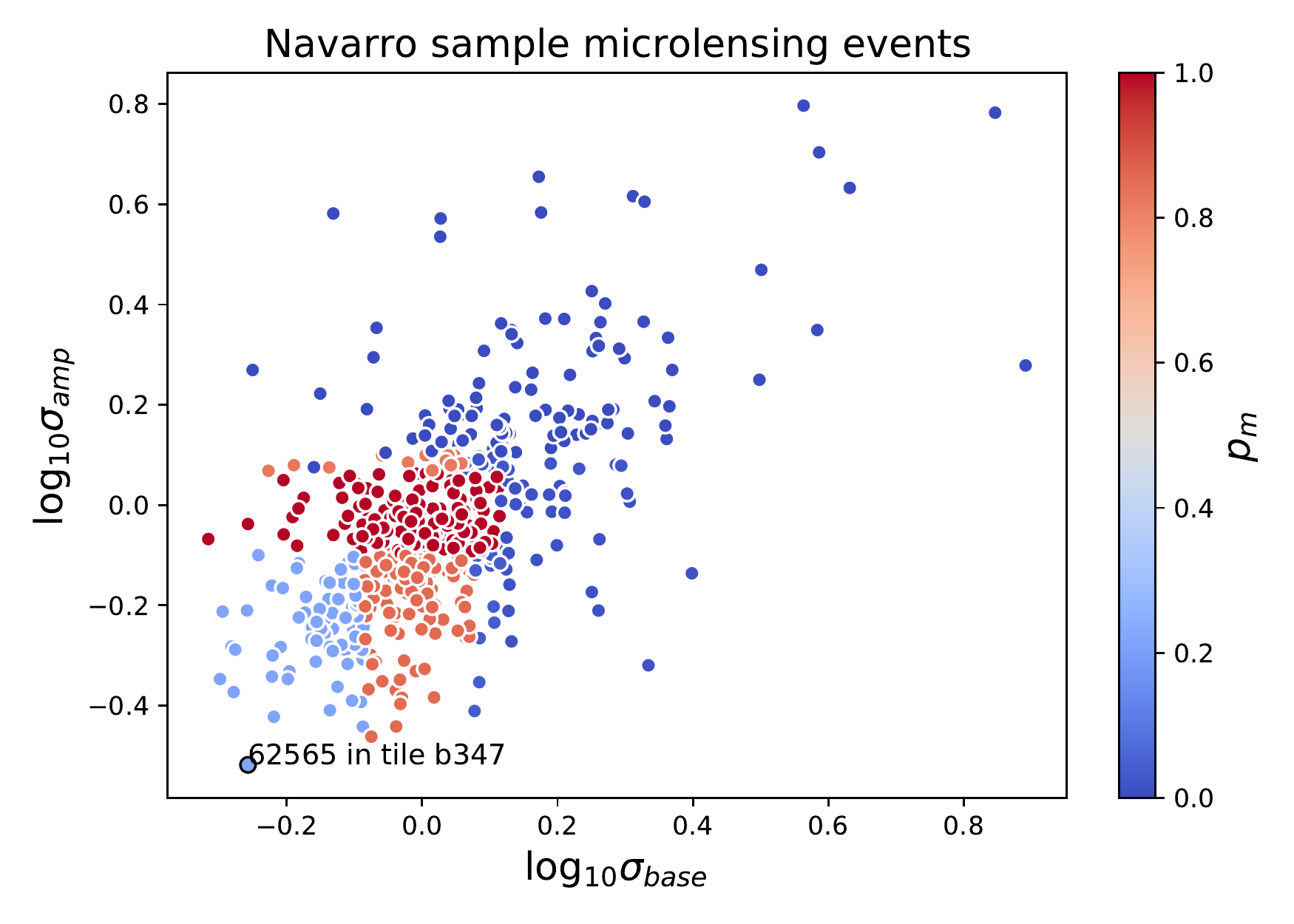}
    \includegraphics[width=\columnwidth]{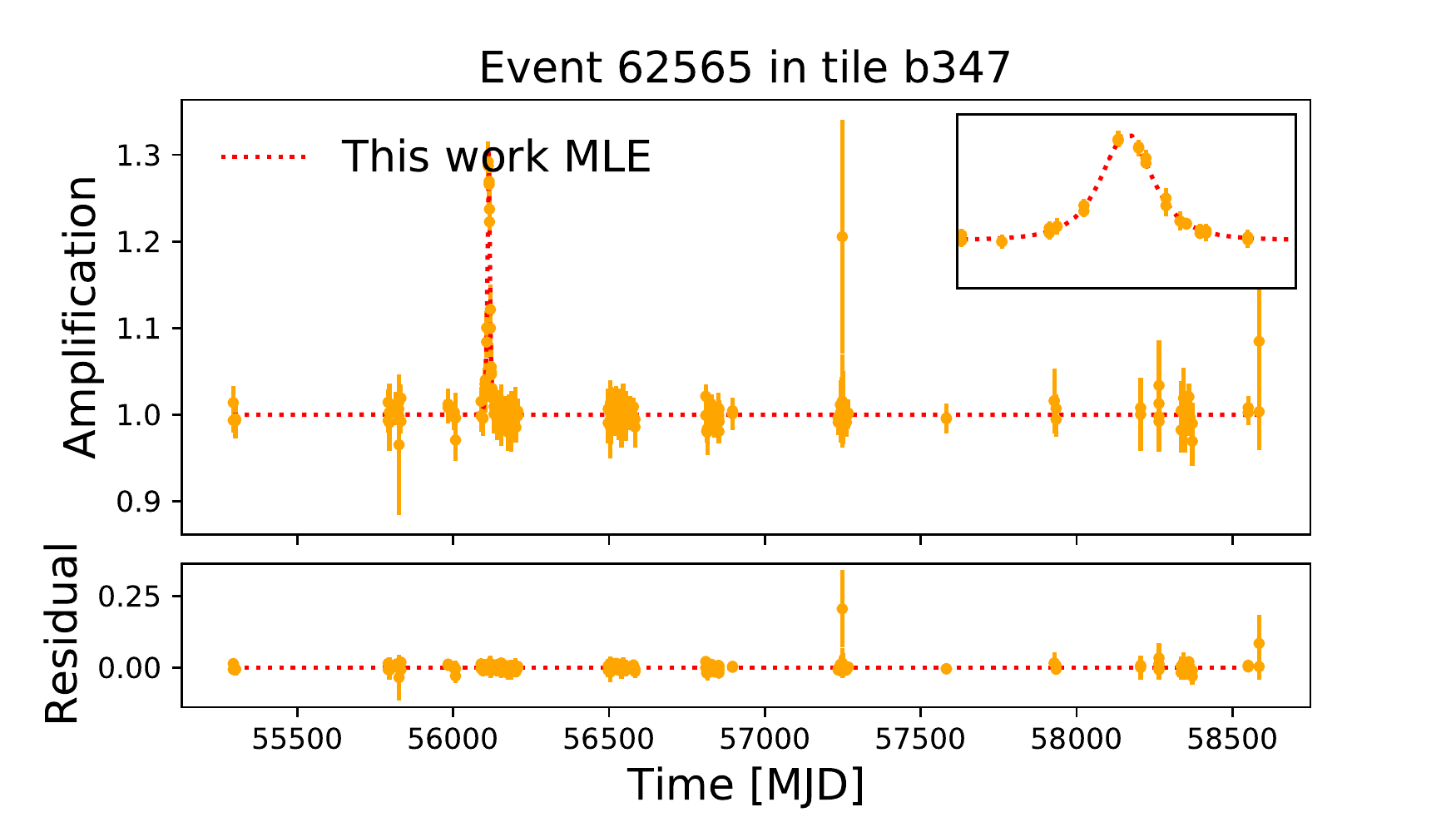}
    \caption{\textbf{Top}: Projection of the feature space in $\sigma_{\text{amp}}-\sigma_{\text{base}}$ for the $526$ previously found Navarro microlensing events that meet our initial quality cuts outlined in table \ref{tab:quality_cuts}. Events are coloured by their Decision \todo{tree microlensing class probability ($p_{m}$)}. \textbf{Bottom}: Light curve of \todo{Navarro ID $62565$ in tile b$347$}, the events position in $\sigma_{\text{amp}}-\sigma_{\text{base}}$ space in indication in the top panel. Inset in the top right is a zoomed-in region of the microlensing amplification. The light curve is plotted in amplification space, which is calculated with respect to the maximum likelihood baseline flux. The maximum likelihood solution from this work shown over the data.}
    \label{fig:navarro_sigma_amp_sigma_base}
\end{figure}

We now focus on the 526 events from the Navarro sample that passed our initial quality cuts. Some of these events did not make our final catalogue. The reason can best be seen by inspecting the decision surface of the classifier in $\sigma_{\text{base}}-\sigma_{\text{amp}}$ space. Fig.~\ref{fig:navarro_sigma_amp_sigma_base} (top panel) shows the 526 events in this space with the decision tree classifier microlensing class probability overlaid. The decision surface shows that events with both high $\sigma_{\text{base}}$ and $\sigma_{\text{amp}}$ are confidently classified as not microlensing (dark blue), which is as expected. In these cases, the light curves have many data points far away from the PSPL maximum likelihood model both on the amplification and in the baseline (taking into account their respective error bars). For events with $\sigma_{\text{amp}}$ and $\sigma_{\text{base}}$ $\approx1$, the classifier confidently labels them as microlensing with high probability (dark red). This reflects the simulated microlensing event data used to train the decision tree classifier. We note that for events with $\sigma_{\text{base}}\approx 1$ but with $\sigma_{\text{amp}}<1.0$, the classifier labels them as microlensing but with less confidence (light red). Promisingly, this indicates that the classifier is using the quality of the maximum likelihood PSPL fit on the microlensing amplification to differentiate between very high probability events (dark red) and high probability events (light red).

However, for events with both low $\sigma_{\text{base}}$ and $\sigma_{\text{amp}}$, the classifier labels them as not microlensing with intermediate probabilities $\approx0.4$ (light blue).  Fig.~\ref{fig:navarro_sigma_amp_sigma_base} (bottom panel) shows the light curve for one of these events (\todo{Navarro ID $62565$ in tile b$347$}). Visual inspection of the light curve clearly identifies \todo{$62565$ in tile b$347$} as microlensing, indicating room for improvement in the classification algorithm. For such events, the PSPL maximum likelihood solution appears to over-fit the light curve data. This is because, both in the baseline and amplification, the data are too tightly distributed around the model given the size of their error bars. If \todo{$62565$ in tile b$347$} is a genuine PSPL microlensing event, this suggests an assumption made during the modeling has been violated. In finding the PSPL maximum likelihood solution (Section~\ref{sec:mle}), simulating the microlensing training data (Section \ref{sec:simulations}), and interpreting the $\sigma_{\text{amp}}$ and $\sigma_{\text{base}}$ features (Section \ref{sec:checking_features}), we assumed independent Gaussian noise with standard deviation of the reported or expected light curve error bars. This is a strong assumption and is unlikely to hold true for all sources in the survey.

For example, the reported error bars could be either over- or under-estimated which could cause $\sigma_{\text{amp}}$ and $\sigma_{\text{base}}$ to be smaller. Moreover, noise in the light curve could be correlated over small timescales \citep[e.g.][]{Go20}, which would change the values of $\sigma_{\text{amp}}$ and $\sigma_{\text{base}}$. These issues could be partly alleviated by relaxing the independent Gaussian noise assumption (e.g. including a free parameter that inflates or shrinks the reported error bars, or inclusion of correlated noise with a Gaussian process). Of course, these additions would come at extra computational cost, which would make the classification algorithm less scalable. 

Finally, for $\sigma_{\text{amp}}$ and $\sigma_{\text{base}}$, we are assuming the maximum likelihood solution is the true model. If this is not representative of the possible values (posterior distribution) of the PSPL parameters, then this assumption will likely cause issues. In cases of microlensing events with low signal to noise, this can cause the posterior in the PSPL to have a large dispersion. In cases with sparse coverage, this can give rise to many plausible solutions and a multi-modal posterior. As we will see in Section~\ref{sec:nested_sampling}, this does turn out to be the case for some events. These issues could be partially rectified by -- instead of using the PSPL maximum likelihood solution residuals -- computing residuals with respect to an approximation to the posterior (e.g. with a Laplace approximation). However, this too would come at the cost of the scalability of the classification algorithm. 

\subsection{Bayesian inference of PSPL parameters}
\label{sec:nested_sampling}

In the Navarro sample, PSPL parameters for each event were obtained using a non-linear least squares fitting routine implemented in the Scipy python package \citep[optimize.curve\_fit;][]{2020SciPy-NMeth}. This routine first finds the MLE of the PSPL model under the same Gaussian noise assumptions as in Section~\ref{sec:mle}, using the gradient based Trust Region Reflective optimization algorithm \citep{branch1999subspace}. In this, the gradients are computed numerically with a finite difference scheme and the algorithm converges onto a local optimum of the log likelihood function to produce a MLE. The curvature of the likelihood function around the MLE is then approximated to first order, permitting a covariance matrix for the maximum likelihood parameters to be computed. The diagonal elements of the covariance matrix contain the standard errors on the PSPL parameters and are reported as one standard deviation errors on the MLE PSPL parameters in the Navarro sample.

In this work, we instead opt for a Bayesian approach to determine the PSPL parameters. We wish to sample the posterior distribution of the PSPL parameters $\boldsymbol{\theta}\equiv[t_0,t_{\rm E},u_0,m_0,f_{\rm bl}]$ for each event. By Bayes theorem, the posterior distribution is given by
\begin{equation}
    p({\boldsymbol\theta}|\mathcal{D}, \text{PSPL}) = \frac{p(\mathcal{D} |{\boldsymbol\theta}, \text{PSPL})p(\boldsymbol{\theta}| \text{PSPL})}{p(\mathcal{D}|\text{PSPL})}
    \label{eq:bayes}
\end{equation}
where
\begin{equation}
    p(\mathcal{D}|\text{PSPL}) = \int_{\Omega_{\boldsymbol{\theta}}}
    p(\mathcal{D}|\boldsymbol{\theta},\text{PSPL})p(\boldsymbol{\theta}| \text{PSPL})d\boldsymbol{\theta}.
    \label{eq:evidence}
\end{equation}
Here, $p(\boldsymbol{\theta}| \text{PSPL})$ is the prior over the PSPL parameters, $p(\mathcal{D} |\boldsymbol{\theta}, \text{PSPL})$ is the likelihood as defined in eqn~(\ref{eq:log_like}), and $p(\mathcal{D}|\text{PSPL})$ is the model evidence which is an integral over all possible values of the PSPL parameters ($\Omega_{\boldsymbol{\theta}}$).  For the fitting, we transform the data into flux space assuming a zero point magnitude of 22.0, meaning we fit a baseline flux $f_{0}$ instead of a baseline magnitude $m_{0}$. We also fit $t_{\rm E}$ and $f_{0}$ in natural logarithm space as these parameters are then restricted to be positive.

\begin{table}
	\centering
	\caption{Description of priors used in the modeling of the PSPL events. Here, $t_{\text{min}}$ and $t_{\text{max}}$ are the minimum and maximum times of each light curve. $F_{\text{med}}$ and $F_{\sigma}$ are the median and standard deviations of the fluxes in the light curve.}
	\begin{tabular}{ll} 
		\hline
		Parameter & Prior distribution \\
		\hline
		$\ln{t_{\rm E}}$ & \text{normal}(mean=3, standard deviation=6) \\
		$u_{0}$  & \text{exponential}(rate=0.5) \\
		$t_{0}$ & \text{uniform}(lower=$t_{\text{min}}$, upper=$t_{\text{max}}$) \\ 
		$\ln F_{\text{base}}$ & \text{normal}(mean=$\ln(F_{\text{med}})$, standard deviation=$\ln(3F_{\sigma})$) \\
		$f_{\rm bl}$ & \text{uniform}(lower=0, upper=1.2) \\
		\hline
	\end{tabular}
\label{tab:priors}
\end{table}

\begin{figure*}
    \centering
    \includegraphics[width=\textwidth]{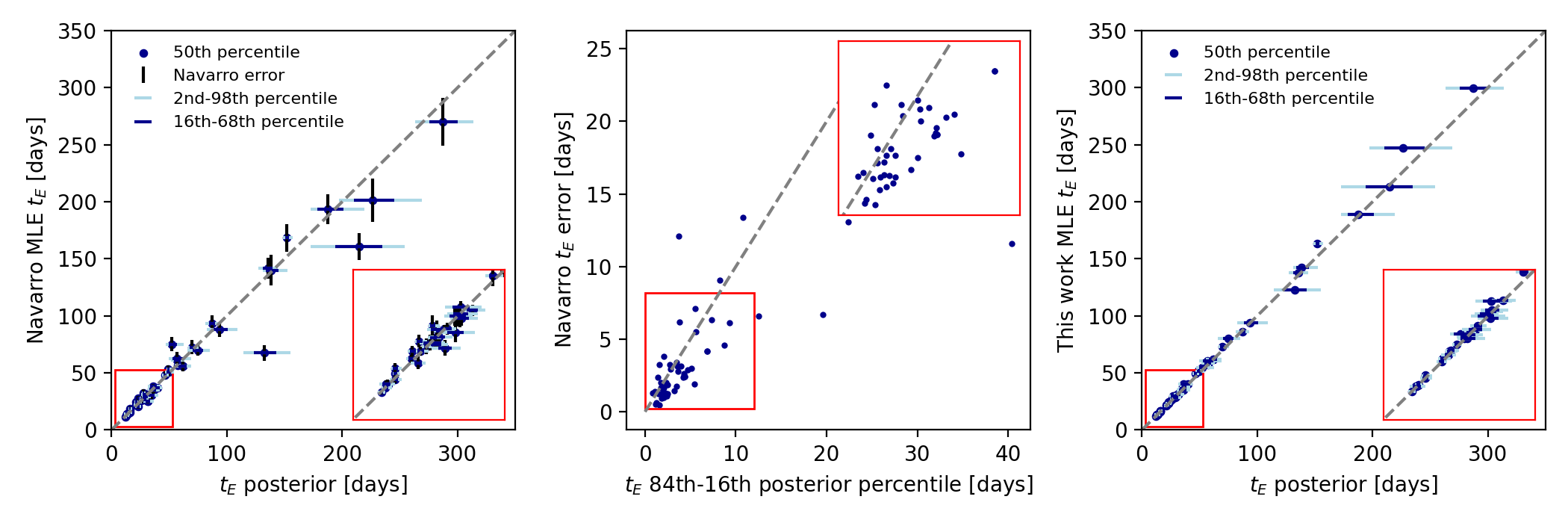}
    \caption{Comparison between the inference on $t_{\rm E}$ between the analysis of the Navarro sample and a Bayesian analysis in this work. All 53 events in the Navarro sample where a $\leq10$\% error on the inferred value of $t_{\rm E}$ is reported by the Navarro analysis are shown. \textbf{Left}: Reported Navarro $t_{\rm E}$ value against Bayesian analysis posterior median from this work. Good agreement for most events is seen. \textbf{Middle}: One standard deviation error from the Navarro analysis against 84th-16th percentile of the posterior from this work. The Navarro analysis tends to underestimate the error for the best constrained events. \textbf{Right}: Maximum likelihood estimate computed in this work (section \ref{sec:mle}) compared with the posterior median. Better agreement is observed compared with the Navarro analysis in the right panel. This is due to the analysis in this work using the global differential evolution optimization algorithm.}
    \label{fig:navarro_comparison}
\end{figure*}

Our strategy is to use weakly informative priors to constrain the model parameters to reasonable, physical regions of the parameter space. The prior used for $t_{0}$ is set dynamically for each event and is uniform between the minimum and maximum times in the light curve. The prior for $\ln f_{0}$ is also set dynamically for each event to be Gaussian with mean of the median flux of the light curve and standard deviation to be three times the standard deviation of the flux in the light curve. The prior for $\ln t_{\rm E}$ is set to be Gaussian with mean 3 ($t_{\rm E}\approx20$ days) and a standard deviation of 6 allowing $t_{\rm E}$ to range from $\approx 10^{-2}-10^{4}$ days. For $u_{0}$, the prior is an exponential with rate = 0.5. This constrains $u_{0}$ to be positive and keeps its value in a reasonable physical range, as we do not expect to have detectable microlensing events with $u_{0}\gg 1$. Finally, the prior used for $f_{\rm bl}$ is uniform with lower and upper bounds to be 0.0 and 1.2 respectively. The upper bound is above the physical limit of 1.0 to allow for a small amount of negative blending. Negative blending or values of $f_{\rm bl}$ just above one are not physical (in such a case the source appears to contribute more than 100 per cent of the flux to the blend), but are possible due to systematic effects in the DoPHOT photometric reduction pipeline \citep{Park2004, Smith2007}. The prior factorises over the PSPL parameters and the details are summarised in Table~\ref{tab:priors}.

To obtain samples from the posterior distribution of the PSPL parameters, we use the dynamic nested sampling algorithm \citep{hi19} implemented by the \textsc{dynesty} Python package \citep{Sp20}. First proposed by \cite{Skilling2004}, nested sampling aims to numerically estimate the model evidence. Nested sampling works by drawing a number of live points from the prior and iteratively removing the point with the lowest likelihood and replacing it with a new point drawn from the prior with a higher likelihood. The procedure allows the prior to be integrated in nested shells of constant likelihood, in turn allowing the model evidence of eqn~(\ref{eq:evidence}) to be estimated. The procedure is repeated until the estimated remaining fractional model evidence drops below some threshold level. As a by-product of this procedure, posterior samples for the model parameters are generated. In the case of dynamic nested sampling, the number of live points are dynamically allocated, which allows a better estimate of the posterior density at a cost of larger error in the model evidence estimate\footnote{See the dynesty documentation (\url{https://dynesty.readthedocs.io/}) and references therein for a detailed explanation of nested sampling.}. In this study, we fit events with 1\,000 initial live points, random walk sampling \citep{sk06}, and multiple bounding ellipsoids. We allocated samples 100 per cent of weight on the posterior, and used a stopping criteria in the remaining fractional evidence of $0.01$.

The main reason for using nested sampling to characterise the posterior distributions as opposed to MCMC based samplers (e.g. ensemble, \cite{Mackey2013F}; or Hamiltonian Monte Carlo, \cite{Go20}) is its ability to sample from complex multi-modal distributions. Due to the sparse and relatively low signal to noise nature of microlensing light curves in the VVV, the posterior distributions for PSPL parameters for some events do turn out to be complex and multi-modal.  As an illustration, Fig.~\ref{fig:mosaic} shows a selection of 24 light curves from our sample of 1\,959 microlensing events. We have chosen them to give an idea of the range of light curves in our catalogue, including examples of high and low amplification events with a variety of noise properties. Overplotted on the $K_{s}$-band data are one hundred posterior realisations, together with the MLE. Notice that, especially for noisy, low amplification events, the posterior realizations yield a much broader range of light curve behaviour than indicated by the maximum likelihood curve.

We now turn to comparing our posterior inferences on $t_{\rm E}$ for each event to the methods used on the Navarro sample. We focus our attention on the 53 events in the Navarro sample with a reported one standard deviation error $<10\%$ of the reported value of $t_{\rm E}$. Fig.~\ref{fig:navarro_comparison} shows the comparison between our posterior inference and those from Navarro. In general, our posterior median for $t_{\rm E}$ is in good agreement with the reported value from the Navarro sample (left panel). However, we find that the fitting routine used in the Navarro sample tends to underestimate the uncertainty in $t_{\rm E}$ when comparing the one standard deviation error with the 84-16th percentile of the posterior for the events with a well constrained $t_{\rm E}$ (middle panel).

Fig.~\ref{fig:navarro_comparison} (right panel) shows the posterior distribution compared to the MLE parameters obtained in this work and described in Section~\ref{sec:mle}. We see better agreement between the posterior median and our MLE parameter values compared with the Navarro sample maximum likelihood solution. This difference is likely due to the fact that the differential evolution optimization algorithm is better able to find the global optimum of the likelihood function compared with gradient based Trust Region Reflective optimization algorithm. 

\begin{figure*}
    \centering
    \includegraphics[width=\textwidth]{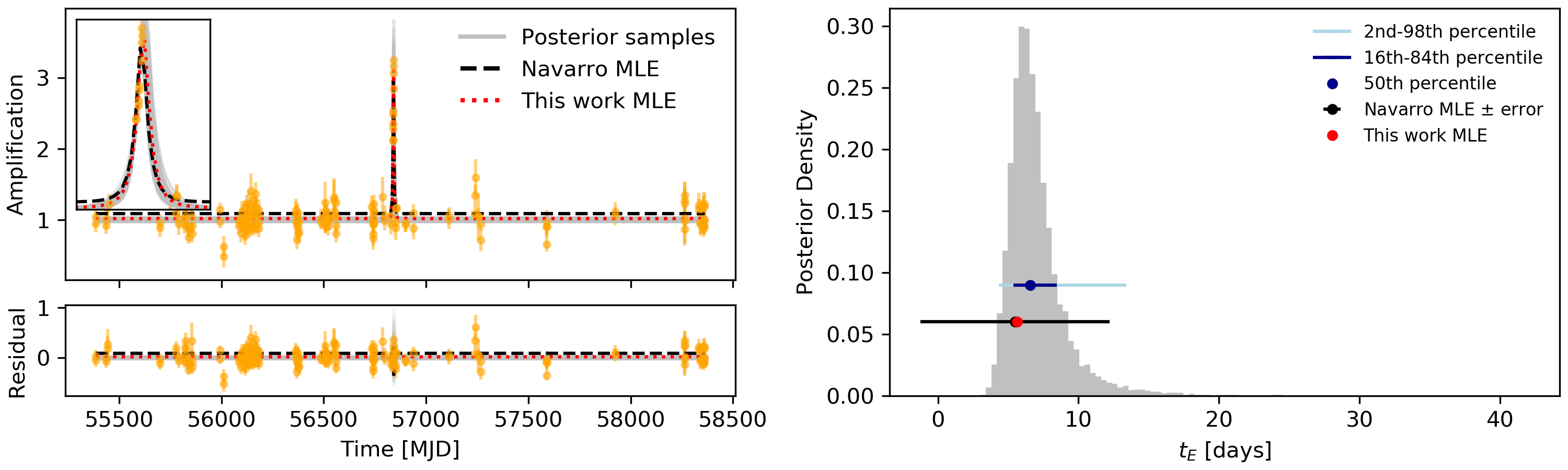}
    \includegraphics[width=\textwidth]{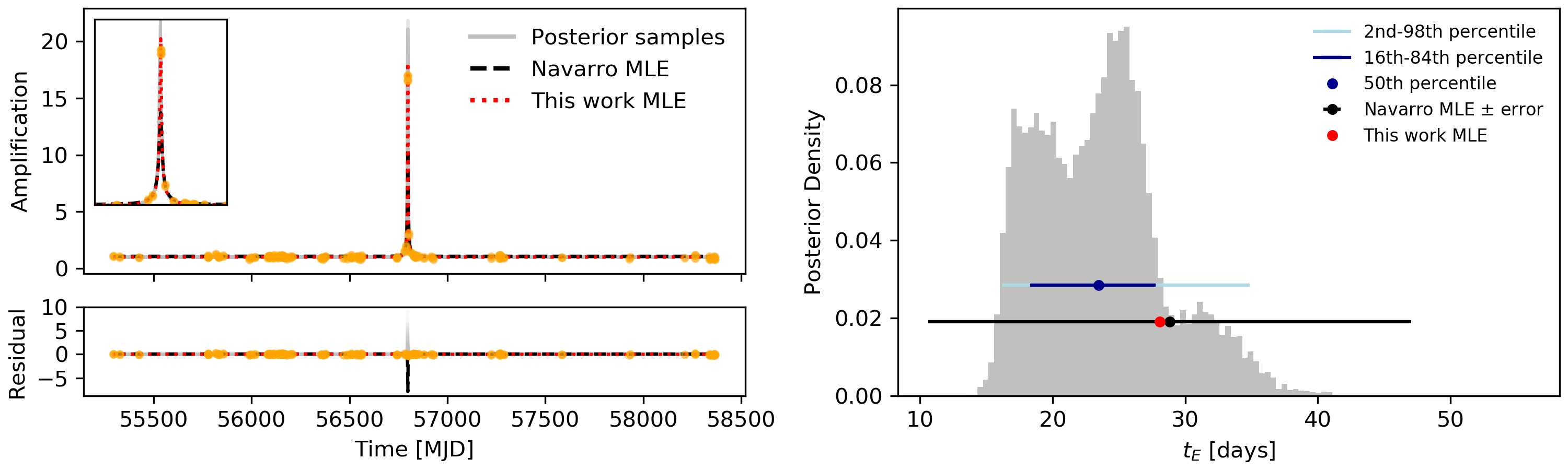}
    \caption{\textbf{Top}: (Left) Light curve of the event with \todo{Navarro ID $79184$ in tile b$305$} with the posterior inference and maximum likelihood estimates (MLE) from the Navarro analysis and this work overlaid. (Right) marginal posterior distribution with percentiles and maximum likelihood solution overlaid. Both MLEs find the mode of the posterior. In this case, Navarro's error estimate fails to capture the asymmetric shape of the posterior. \textbf{Bottom}: Same as the top row, but for the event with \todo{Navarro ID $83519$ in tile b$337$}. In this case, both MLEs fail to capture the multi-modal nature of the posterior.}
    \label{fig:inference_over_data}
\end{figure*}

Fig.~\ref{fig:inference_over_data} shows an example of two complex posterior distributions for $t_{\rm E}$ which are poorly characterised by the maximum likelihood solution. The top row shows the event with \todo{Navarro ID $79184$ in tile b$305$}. In this case, both the maximum likelihood methods from this work and Navarro's analysis find the mode of the posterior. However, the Navarro error estimate overestimates the spread of the posterior. Moreover, the Navarro error estimate fails to capture the asymmetric nature of the posterior distribution which has a heavy right tail. The bottom row of Fig.~\ref{fig:inference_over_data} shows the event with \todo{Navarro ID $83519$ in tile b$337$}. In this \todo{event}, the posterior is multi-model in $t_{\rm E}$, and both MLEs fail to precisely find the major mode. Although the Navarro error estimate covers most of the posterior distribution, it does not capture its complex shape. 

We note that in many cases offsets in the Navarro baseline magnitude and the baseline from our reduction of the VVV data are present. This is likely due to differences in the photometric processing of the VVV images, specifically their use of alternative profile fit photometry software and photometric calibration methods. We conclude that the sparse and low signal to noise ratio nature of many microlensing events in the VVV necessitate a full Bayesian analysis to fully characterise the inferences on $t_{\rm E}$.

\section{Inferences on the Einstein timescale}
\label{sec:tE_inference}

For the 1959 microlensing events in our catalogue, we fit the PSPL model by the Bayesian method described in Section \ref{sec:nested_sampling}, and obtain posterior distributions for the Einstein timescale $t_{\rm E}$. In Appendix A, Table \ref{tab:event_inference} contains the 16th, 50th, and 84th posterior percentiles of $t_{\rm E}$ for each event along with the same percentiles for all of the other PSPL parameters. 

\begin{figure}
    \centering
    \includegraphics[width=\columnwidth]{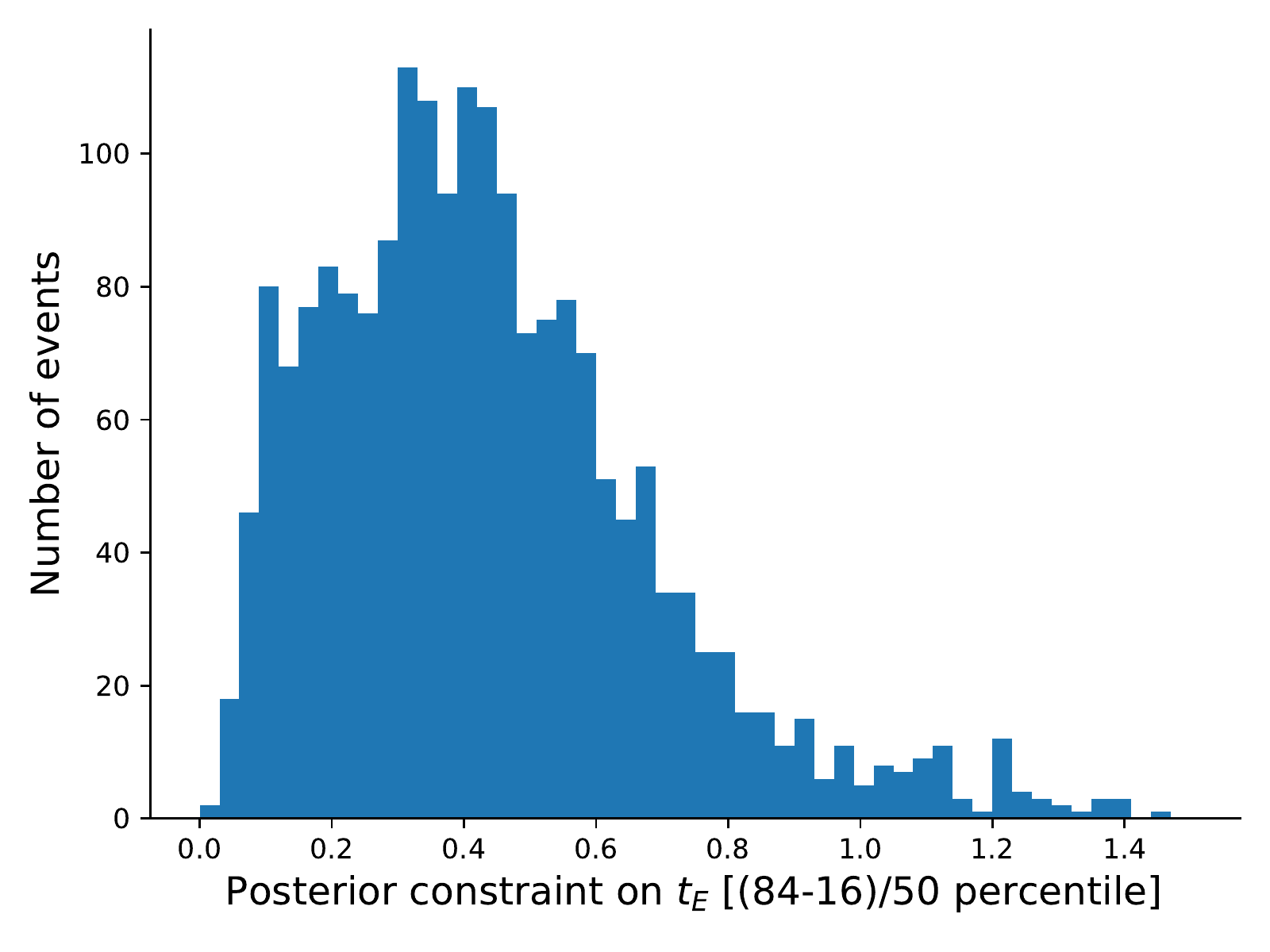}
    \includegraphics[width=\columnwidth]{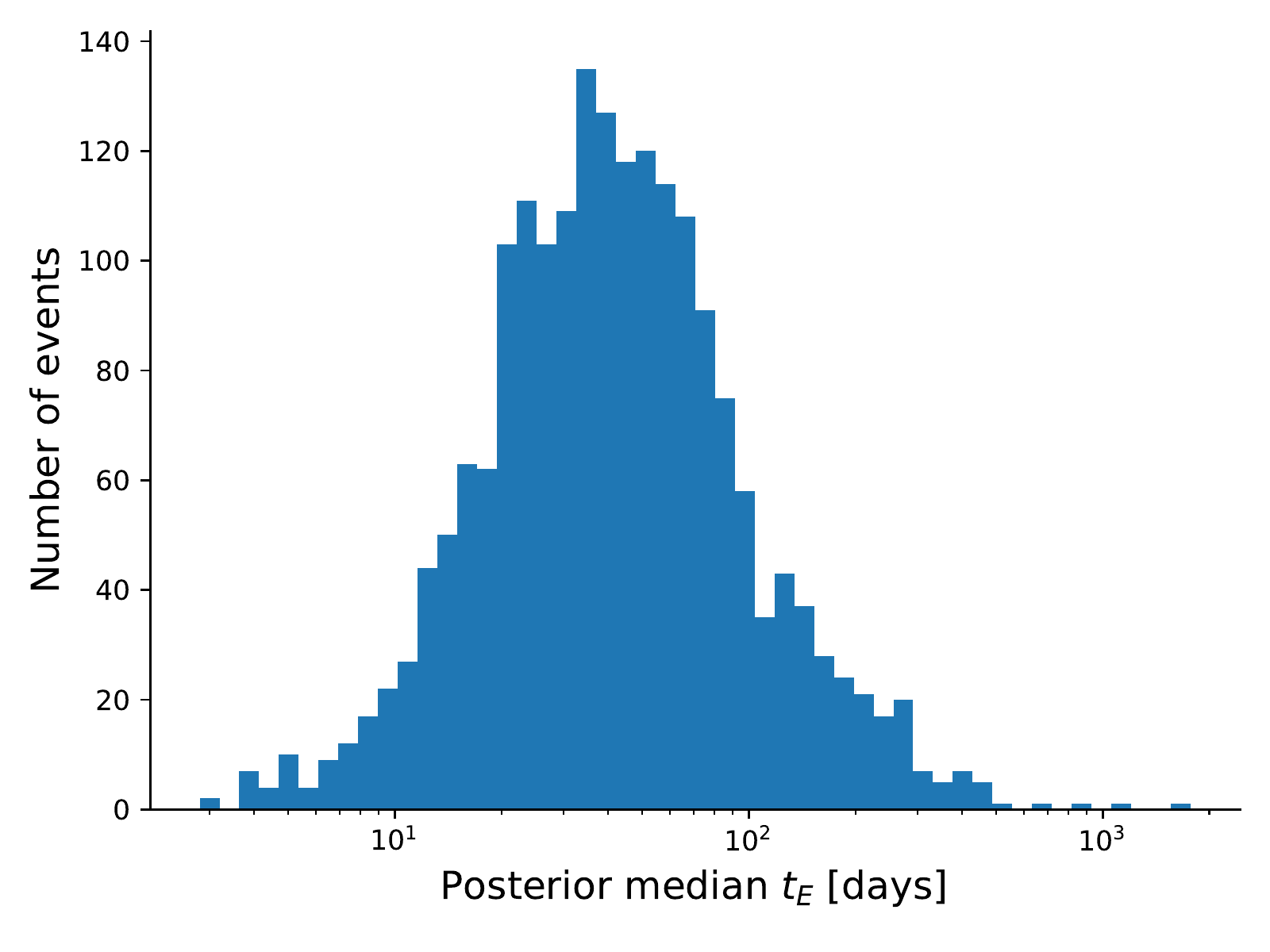}
    \caption{\textbf{Top}: Distribution of the posterior constraint provided on $t_{E}$ for all 1959 microlensing events found in this work. Typically events have $t_{E}$ constrained within $\sim$40\%. \textbf{Bottom}: Distribution of the median values of the posterior on $t_{E}$ for all the microlensing events.}
    \label{fig:tE_constraint}
\end{figure}

The top panel of Fig.~\ref{fig:tE_constraint} shows the distribution of constraints on $t_{\rm E}$ obtained by the Bayesian analysis of the events. We can see that typically $t_{\rm E}$ is only constrained within $\sim40$ per cent. These constraints are somewhat poorer than those typically achieved by dedicated microlensing surveys such as OGLE, with better coverage of events and higher observing cadences \todo{\citep[e.g. see Figure 2 in ][]{Mr19}}.

The bottom panel of Fig.~\ref{fig:tE_constraint} shows the distribution of the median of the posterior of $t_{\rm E}$. It shows median values of $t_{\rm E}$ ranging from $10^{0}-10^{3}$ days and is peaked at $\sim30$ days. One may be tempted to convolve this histogram in its raw form with detection efficiency \citep[as is commonly done in microlensing analyses e.g.][]{Mr20new} in order to make inferences about the timescale $t_{\rm E}$ distribution over all events. This method however neglects the uncertainty on the $t_{\rm E}$ value for each event, which is not negligible for the sample of microlensing events found in this work. 

A method that allows the population level distributions to be inferred from a set of noisy objects is presented in \citet{Hogg2010} and \citet{Foreman-Mackey2014}. It was applied while inferring the population $t_{\rm E}$ distribution of OGLE microlensing events by \cite{Go20}. By modeling in this way, and accounting for correlated noise and parallax, \cite{Go20} showed this can have a large impact on the high end of the inferred $t_{\rm E}$ population distribution when compared to the simple histogram approach. Given the large uncertainty in each event $t_{\rm E}$ in the sample presented here, robust inference of population level distributions will likely be best achieved by careful application of the methods described by \citet{Hogg2010} and \citet{Foreman-Mackey2014}. We therefore defer this analysis for future work.

\section{Conclusions}

We have performed the first search over the full VVV survey area of the inner Galaxy and provided a sample of 1\,959 microlensing events. We achieved this using a comparatively simple and interpretable machine learning algorithm with very few features. Although applications of machine learning to microlensing have been studied before~\citep{Wy15,Go19}, this is the first successful attempt to implement an algorithm on the full \todo{VVV} dataset of \todo{$\sim700$ million lightcurves.}

We used a decision tree classifier, which is a non-parametric supervised learning algorithm. Given a set of features, the classifier splits the data into two partitions, here microlensing and non-microlensing. This happens iteratively, as the data are submitted to more and more tests and split further and further on each of the branches of the decision tree, \todo{producing a fast and scalable classifier}. Unlike earlier microlensing classifiers~\citep{Wy15,Go19}, we used just 4 features -- specifically, standardised residuals between the data and the MLE of the point source-point lens (PSPL) model both (i) inside and (ii) outside two Einstein times of maximum amplitude, (iii) the change in the Akaike information criterion between PSPL and constant light curve models and (iv) the von Neumann ratio. The classifier achieves 97 per cent accuracy on identifying microlensing events in training data comprised of random VVV light curves and simulated microlensing and cataclysmic variable events.

Even though our training set contains $\sim 10^4$ examples, there are still some anomalous, very noisy or unusual VVV light curves that are not recognised. These can end up being misclassified and have to be removed by hand. So, the very final stage requires human intervention. This though is at a manageable level of inspecting $\sim10^4$ light curves from the original starting sample of $\sim 10^9$. So, the classifier has made the task of analyzing the entire VVV light curve database manageable. The final catalogue of 1\,959 microlensing events, together with their positions and inferences on their timescales, impact parameters and blending fractions, are available in electronic form in Table~\ref{tab:event_inference}. Some sample light curves are displayed in Fig.~\ref{fig:mosaic}, together with maximum likelihood and Bayesian modelling. 

Parts of the VVV footprint have previously been scrutinised for microlensing by \citet{Navarro2018,NavarroLatitude,NavarroForesaken,NavarroFardisk}. They used a conventional mixture of parameter cuts to remove variable stars and other contaminants, combined with visual assessment. Our sample is less complete compared to the previous Navarro et al. searches in the common regions of sky, but it extends over the whole VVV survey area and possesses a \todo{better} characterised selection function. \todo{However, we do acknowledge that in the final visual classification step, we do potentially introduce hard to quantify error into our selection function. This is partially mitigated by averaging visual classification of different people for each event, and by precise characterisation of the decision tree selected sample, of which the visually confirmed events are a subset.} 

For our events, we provide spatially resolved classifier efficiency maps as a function of the Einstein time $t_{\rm E}$. This is vital for population studies. For example, these efficiency maps shows large features in $t_{\rm E}$ due to variable survey observing strategies, and are important to take into account when doing spatially resolved population studies. The latitude or longitude dependence of the optical depth or timescale distribution contains information on the lens and source population~\citep{EB}, but it is not possible to extricate without understanding the variation of efficiency on the sky~\citep[cf][]{NavarroLatitude}.
  
Due to the relatively sparse and noisy VVV light curve data as compared with dedicated microlensing surveys like OGLE, we implemented a Bayesian scheme to determine the parameters of our 1\,959 events. We find that the posterior distributions for $t_{\rm E}$ are often complex. We show that maximum likelihood approaches used in previous analyses do not capture this complexity and therefore fail to entirely characterise the uncertainty in $t_{\rm E}$. Only for the well constrained events ($<10\%$ error on $t_{\rm E}$) do we find good agreement between our Bayesian approach and MLEs. This has implications for forthcoming surveys that are not dedicated to microlensing and so have sporadic sampling, such as the {\it Legacy Survey of Space and Time} (LSST) at the Vera Rubin Observatory~\citep{Sa19} or the {\it Zwicky Transient Facility} (ZTF) at Palomar Observatory~\citep{Me20}. Also, for many events in our sample, there is typically a $\sim40$ per cent constraint on $t_{\rm E}$. This necessitates a Bayesian approach to population inference to handle uncertainty \citep[e.g.][]{Go20}, rather than previous approaches of histograms which ignore the uncertainty in each $t_{\rm E}$. For the VVV survey, $t_{\rm E}$ is typically not well enough constrained to study distributions of timescales without allowing for this.
 
In our final step involving visual inspection, every light curve was examined by at least $3$ of the $4$ authors. The consistency of any individual is often good, in the sense that when provided with the same light curve again, the same decision as to microlensing or non-microlensing is made. However, from individual to individual, there is considerable variation as to the classification, especially if the light curve is noisy or sparsely sampled. We dealt with this by averaging over the scores of all the individuals participating in our visual assessment. However, visual inspection is not completely reliable as a final step. In ideal circumstances, we would prefer a fully automated classification algorithm. We suspect that the way to improve ours is to model and simulate the noise properties of the data better.

There are a number of future directions for the work in this paper. First, we plan to carry out a statistical analyses of the longitude and latitude dependence of properties of our sample of VVV microlensing events. This has already been examined over a smaller portion of the VVV footprint in \citet{Navarro2018,NavarroLatitude}. The advantage of revisiting the problem with the new sample is that our efficiency maps enable us to remove spatially dependent artefacts associated with properties of the survey or extinction. With these effects untangled, we can reconstruct the distribution of timescales $t_{\rm E}$ as a function of position. In our sample, the timescale distribution between $t_{\rm E} = 30$ days and $100$ days drops by $\sim$ 50\% in the bulge fields, but appears to remain fairly constant in the disc. Is the change in recovery efficiency large enough to explain this or is this a real feature? An answer to this question will give us clues as to the properties of the lens and source populations.

Secondly, the number and properties of microlensing events depend on the distribution and kinematics of stars and compact objects along the line of sight. Precise measurements of the microlensing optical depth and the event rate toward the Galactic bulge enable tests of competing models of the inner Milky Way, including the orientation and mass of its bar. Such calculations are only possible because of our accurate characterisation of the efficiency of the classification algorithm for events. The most recent calculations by OGLE are based on $\sim 8000$ events with optical light curves~\citep{Mr19}. However, the infrared bands of the VVV survey are better able to penetrate the extinction close to the Galactic plane. Our microlensing sample is a complementary probe to OGLE in the inner Galaxy.  

Thirdly, anomalous microlensing events can show deviations from the 
PSPL model. Of course, it is the important category of binary lenses that shows the most substantial morphological variety. The light curves can exhibit multiple peaks and caustic crossings and can be very different to the familiar time-symmetric PSPL light curve~\citep[e.g.,][]{Ma91,Di97}. There are also less drastic changes caused by parallactic or finite source-size effects for single lenses. Our classifier is not trained to identify such light curves and we expect most in the VVV survey will have been missed. The problem of extending machine learning methods to identify anomalous events is both interesting and challenging~\citep{Mr2020, Khakpash2021}. In the wider context, the VVV survey is already proving a useful testing ground for automatic classification of all manner of variable phenomena, such as RR Lyrae~\citep{Ca20} and other types of variable star~\citep{Me18}. We expect the importance of this activity to increase, as we prepare for larger surveys and larger telescopes to map out the terrain of time-domain astronomy. 

\section*{Acknowledgements}

AH acknowledges support from the summer studentship program at the IoA, Cambridge, run by Vasily Belokurov and Matt Auger. PM acknowledges STFC for studentship funding. This work has made use of the University of Hertfordshire's high-performance computing facility. The authors would like to thank Gabriela Navarro \todo{and members of the Cambridge streams groups} for useful discussions. \todo{The (anonymous) referee is thanked for a helpful and thoughtful report which improved the clarity of the paper.}

\section*{Data Availability}

The electronic table of microlensing events is published with this article as supplementary material. All other codes and products  (such as the efficiency maps) are available from the authors on request.



\bibliographystyle{mnras}
\bibliography{ref}



\appendix

\section{The Catalogue}

\begin{figure*}
    \centering
    \includegraphics[width=\textwidth]{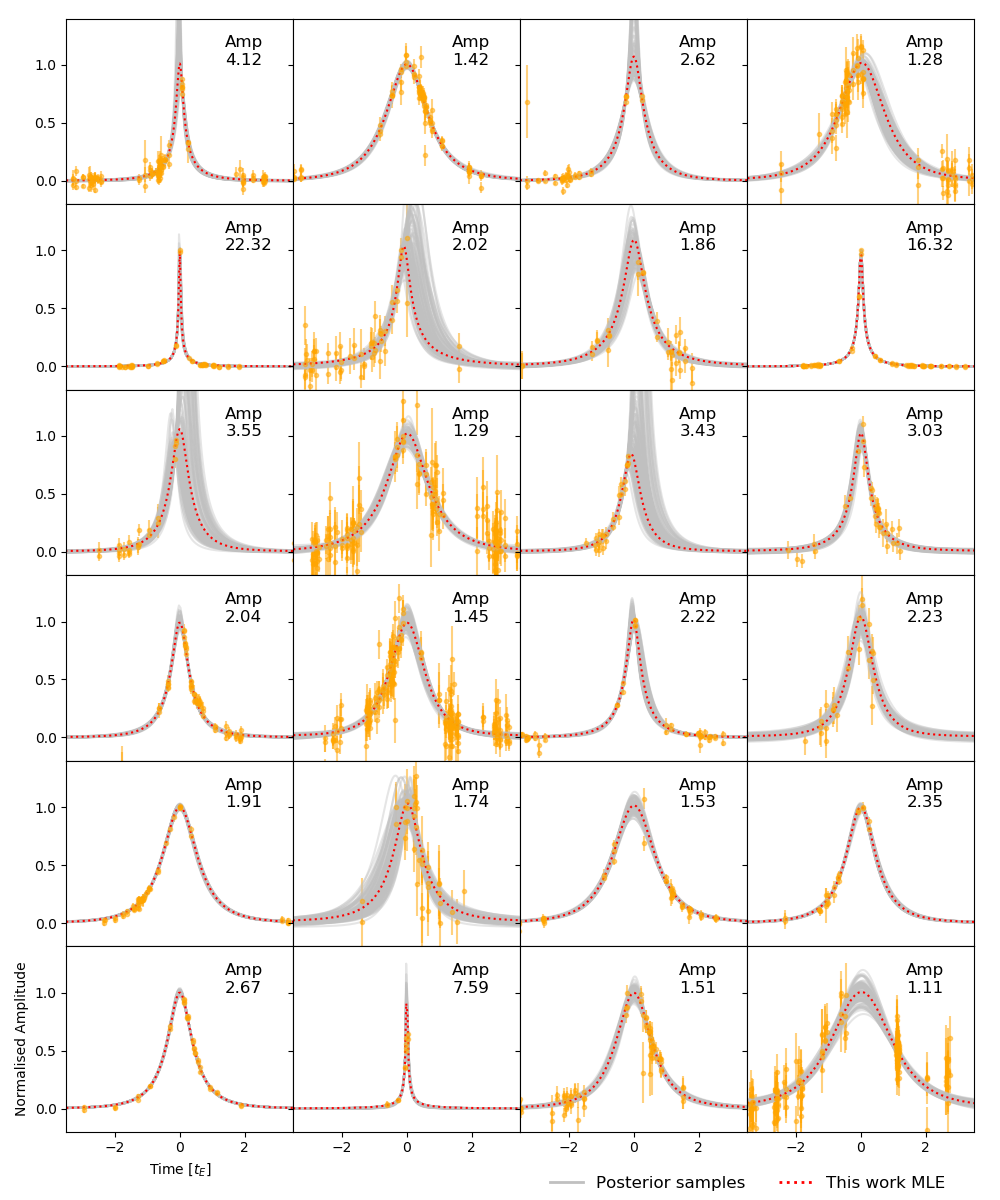}
    \caption{Selection of found microlensing events with one hundred posterior realisations and the maximum likelihood estimate (MLE) overlaid on the $K_{s}$-band light curve data. Each event is plotted in normalised amplitude and $t_{\rm E}$ space, to allow them all to be plotted on the same scale. For each event the flux is offset by the posterior median baseline flux and normalised by the posterior median of the flux at maximum amplification. An estimate of the maximum amplification is shown as text in each plot. This maximum amplification is calculated by dividing the posterior median maximum flux by the posterior median baseline flux.}
    \label{fig:mosaic}
\end{figure*}

We provide here some details of the catalogue of 1\,959 microlensing events. In Fig.~\ref{fig:mosaic}, a sample of 24 light curves is shown, together with realisations from the posteriors from our Bayesian fitting. The amplitude of the event is indicated in the top right of each panel. The mosaic shows a mix of clear-cut, high amplitude events and noisier, low amplitude events. Table~\ref{tab:event_inference} gives the details on each event, labelled in the standard format VVV-date-DSC-Number or VVV-date-BLG-Number. The data gives the year in which the maximum of the event occurs, whilst DSC or BLG refers to the VVV field in which the event is located, according to whether the absolute value of the Galactic longitude $|\ell|$ is greater or less than $10^\circ$. Of course, this does not necessarily mean that the source or lens lie in the disc or bulge. The table gives the posterior inferences on all our events from the Bayesian modelling.

\begingroup
\renewcommand{\arraystretch}{1.4}
\begin{table*}
    \centering
    \caption{Found microlensing events. Posterior inferences on the PSPL microlensing model parameters are reported for each event. Median posterior values with $84$\todo{th}-$50$\todo{th} percentile indicated as a superscript and $16$\todo{th}-$50$\todo{th} percentile indicated as a subscript are shown. \todo{$30$} of the found microlensing events are shown here, the full machine-readable table of $1959$ events can be found in the online supplementary material associated with this paper. Right ascension (ra) and declination (dec) are from preliminary results of the VVV Infrared Astrometric Catalogue (VIRAC) version $2$ \citep[see][for version 1]{virac}. \todo{Positions} are on the International Celestial Reference Frame at epoch $2015.5$ Julian years, and were calculated using references stars from Gaia Data Release 2 \citep{GaiaDR2}. Events are ordered in the table by their posterior median time of closest approach - $t_{0}$ The year of the closest approach is indicated in the event ID, along with location of the event BLG (VVV Galactic bulge field, $|\ell| < 10^\circ$) or DSC (VVV Galactic disc field, $|\ell| > 10^\circ$). \todo{DET column indicates if the event was previously found in the Navarro sample \citep[-N;][]{NavarroLatitude, NavarroForesaken, NavarroFardisk} or by OGLE \citep[O-;][]{Mr19, Mr20new}, or both (ON), or neither (--)}.}
 
    \begin{tabular}{l|l|l|l|l|l|l|l|l|l|}
    \hline
    event ID & ra [deg] & dec [deg] & $m_{0}$ [K$_{s}$ mag] & $u_{0}$ & $t_{\rm E}$ [days] & $t_{0}$ [MJD] & $f_{\rm bl}$ &  DET \\ 
    \hline
VVV-2010-DSC-0001 & 196.9165844 & -62.5247743 & $15.498_{-0.005}^{+0.005}$ & $0.460_{-0.203}^{+0.189}$ & $17.017_{-2.936}^{+5.784}$ & $55390.005_{-1.527}^{+1.588}$ & $0.562_{-0.303}^{+0.382}$ & --\\
VVV-2011-DSC-0001 & 202.0983177 & -63.7661068 & $14.315_{-0.002}^{+0.002}$ & $0.650_{-0.420}^{+0.381}$ & $88.079_{-24.245}^{+37.600}$ & $55669.154_{-30.571}^{+36.826}$ & $0.553_{-0.361}^{+0.384}$ & --\\
VVV-2011-BLG-0002 & 269.8242208 & -29.8006948 & $13.207_{-0.001}^{+0.001}$ & $0.578_{-0.171}^{+0.149}$ & $25.436_{-6.991}^{+6.599}$ & $55693.519_{-0.857}^{+0.659}$ & $0.721_{-0.295}^{+0.316}$ & --\\
VVV-2011-DSC-0003 & 235.8103905 & -56.1199059 & $13.199_{-0.002}^{+0.002}$ & $0.600_{-0.202}^{+0.156}$ & $83.372_{-9.176}^{+16.980}$ & $55707.894_{-3.319}^{+3.341}$ & $0.679_{-0.289}^{+0.312}$ & --\\
VVV-2011-DSC-0004 & 185.0334915 & -62.6911158 & $16.833_{-0.015}^{+0.015}$ & $0.049_{-0.022}^{+0.030}$ & $76.420_{-17.447}^{+34.614}$ & $55728.591_{-0.944}^{+0.885}$ & $0.587_{-0.233}^{+0.273}$ & --\\
VVV-2011-DSC-0005 & 253.8321695 & -44.6821945 & $16.102_{-0.005}^{+0.006}$ & $0.072_{-0.011}^{+0.009}$ & $71.204_{-4.313}^{+7.507}$ & $55730.122_{-1.015}^{+0.920}$ & $1.016_{-0.158}^{+0.116}$ & --\\
VVV-2011-DSC-0006 & 239.1345956 & -52.5095320 & $14.511_{-0.002}^{+0.002}$ & $0.671_{-0.046}^{+0.027}$ & $121.279_{-4.653}^{+6.012}$ & $55738.387_{-1.600}^{+1.555}$ & $1.094_{-0.120}^{+0.072}$ & --\\
VVV-2011-DSC-0007 & 239.9614121 & -52.0903176 & $14.435_{-0.002}^{+0.002}$ & $1.088_{-0.390}^{+0.296}$ & $39.967_{-6.758}^{+12.886}$ & $55747.390_{-3.877}^{+4.570}$ & $0.560_{-0.313}^{+0.372}$ & --\\
VVV-2011-DSC-0008 & 192.2394790 & -60.8963478 & $14.232_{-0.002}^{+0.002}$ & $0.697_{-0.158}^{+0.099}$ & $76.300_{-15.578}^{+17.254}$ & $55749.654_{-10.157}^{+7.558}$ & $0.855_{-0.308}^{+0.224}$ & --\\
VVV-2011-DSC-0009 & 220.5198779 & -60.8167949 & $13.561_{-0.001}^{+0.001}$ & $0.911_{-0.372}^{+0.304}$ & $96.459_{-17.755}^{+38.553}$ & $55751.063_{-6.786}^{+7.254}$ & $0.486_{-0.297}^{+0.414}$ & --\\
VVV-2011-DSC-0010 & 213.9510808 & -60.0892905 & $14.546_{-0.002}^{+0.002}$ & $0.634_{-0.361}^{+0.375}$ & $43.367_{-10.625}^{+15.436}$ & $55751.094_{-14.212}^{+16.213}$ & $0.550_{-0.322}^{+0.418}$ & --\\
VVV-2011-BLG-0011 & 266.6498266 & -26.8943508 & $15.938_{-0.007}^{+0.007}$ & $0.328_{-0.195}^{+0.194}$ & $52.183_{-14.737}^{+17.330}$ & $55754.642_{-13.258}^{+14.370}$ & $0.744_{-0.331}^{+0.303}$ & --\\
VVV-2011-DSC-0012 & 248.7860941 & -49.6747714 & $14.483_{-0.002}^{+0.002}$ & $0.116_{-0.070}^{+0.076}$ & $54.528_{-2.145}^{+3.012}$ & $55762.348_{-0.355}^{+0.351}$ & $1.026_{-0.111}^{+0.103}$ & --\\
VVV-2011-DSC-0013 & 239.0282511 & -54.3301066 & $16.188_{-0.007}^{+0.007}$ & $0.561_{-0.261}^{+0.233}$ & $75.030_{-29.098}^{+33.451}$ & $55763.114_{-31.772}^{+21.005}$ & $0.651_{-0.322}^{+0.354}$ & --\\
VVV-2011-DSC-0014 & 213.9539459 & -61.4693055 & $16.576_{-0.011}^{+0.011}$ & $0.405_{-0.159}^{+0.100}$ & $33.291_{-7.119}^{+13.663}$ & $55778.350_{-4.816}^{+2.253}$ & $0.795_{-0.374}^{+0.275}$ & --\\
VVV-2011-DSC-0015 & 212.2493240 & -60.8625351 & $15.698_{-0.006}^{+0.006}$ & $0.097_{-0.010}^{+0.009}$ & $141.123_{-9.853}^{+13.254}$ & $55781.122_{-0.450}^{+0.427}$ & $1.018_{-0.114}^{+0.102}$ & --\\
VVV-2011-DSC-0016 & 258.0445663 & -41.1322726 & $12.890_{-0.001}^{+0.001}$ & $0.710_{-0.196}^{+0.157}$ & $29.383_{-3.125}^{+5.167}$ & $55781.645_{-3.148}^{+3.870}$ & $0.750_{-0.294}^{+0.277}$ & --\\
VVV-2011-DSC-0017 & 210.4457880 & -62.7339161 & $14.303_{-0.002}^{+0.002}$ & $0.681_{-0.229}^{+0.153}$ & $43.625_{-11.417}^{+17.472}$ & $55782.296_{-4.097}^{+2.010}$ & $0.715_{-0.345}^{+0.322}$ & --\\
VVV-2011-DSC-0018 & 233.8390857 & -57.1626459 & $14.140_{-0.002}^{+0.002}$ & $1.214_{-0.415}^{+0.289}$ & $36.168_{-9.231}^{+13.063}$ & $55784.180_{-9.424}^{+7.264}$ & $0.549_{-0.319}^{+0.370}$ & --\\
VVV-2011-DSC-0019 & 216.2642336 & -61.2198330 & $13.577_{-0.002}^{+0.002}$ & $0.955_{-0.409}^{+0.271}$ & $24.506_{-4.795}^{+12.361}$ & $55792.667_{-1.061}^{+0.987}$ & $0.542_{-0.344}^{+0.392}$ & --\\
VVV-2011-BLG-0020 & 270.0228100 & -30.9901170 & $16.430_{-0.008}^{+0.008}$ & $0.265_{-0.155}^{+0.136}$ & $36.982_{-6.602}^{+13.506}$ & $55794.096_{-0.673}^{+0.725}$ & $0.685_{-0.303}^{+0.334}$ & O-\\
VVV-2011-BLG-0021 & 264.7245092 & -32.7168752 & $13.216_{-0.001}^{+0.001}$ & $0.090_{-0.054}^{+0.062}$ & $37.537_{-1.870}^{+2.823}$ & $55799.874_{-0.727}^{+0.445}$ & $1.042_{-0.135}^{+0.100}$ & --\\
VVV-2011-BLG-0022 & 267.0638336 & -29.4166830 & $14.704_{-0.003}^{+0.003}$ & $0.494_{-0.291}^{+0.288}$ & $27.494_{-7.125}^{+22.131}$ & $55802.409_{-2.043}^{+2.205}$ & $0.383_{-0.263}^{+0.441}$ & --\\
VVV-2011-BLG-0023 & 267.7925075 & -28.7717205 & $13.379_{-0.001}^{+0.001}$ & $0.309_{-0.046}^{+0.048}$ & $35.735_{-3.163}^{+3.976}$ & $55803.509_{-0.470}^{+0.503}$ & $0.777_{-0.148}^{+0.171}$ & --\\
VVV-2011-DSC-0024 & 241.2023854 & -51.2681987 & $14.837_{-0.003}^{+0.003}$ & $0.024_{-0.007}^{+0.010}$ & $415.522_{-111.665}^{+170.634}$ & $55805.336_{-0.875}^{+0.855}$ & $0.025_{-0.008}^{+0.010}$ & --\\
VVV-2011-DSC-0025 & 236.2966835 & -55.1747467 & $13.601_{-0.002}^{+0.002}$ & $0.954_{-0.586}^{+0.334}$ & $11.992_{-2.387}^{+11.220}$ & $55808.195_{-0.897}^{+0.858}$ & $0.447_{-0.356}^{+0.422}$ & --\\
VVV-2011-BLG-0026 & 267.5950792 & -29.4057786 & $14.525_{-0.003}^{+0.003}$ & $0.197_{-0.072}^{+0.049}$ & $30.126_{-2.997}^{+6.343}$ & $55808.953_{-0.517}^{+0.628}$ & $0.857_{-0.280}^{+0.222}$ & -N\\
VVV-2011-DSC-0027 & 227.9546500 & -57.5842466 & $15.736_{-0.013}^{+0.013}$ & $0.191_{-0.119}^{+0.154}$ & $7.421_{-2.964}^{+3.589}$ & $55809.073_{-0.965}^{+1.079}$ & $0.722_{-0.339}^{+0.297}$ & --\\
VVV-2011-BLG-0028 & 263.5326360 & -31.6724423 & $14.661_{-0.002}^{+0.002}$ & $0.251_{-0.125}^{+0.098}$ & $7.947_{-0.783}^{+1.161}$ & $55811.805_{-0.112}^{+0.125}$ & $0.810_{-0.217}^{+0.242}$ & --\\
VVV-2011-BLG-0029 & 261.7799914 & -29.1655239 & $16.376_{-0.007}^{+0.007}$ & $0.312_{-0.168}^{+0.147}$ & $13.764_{-2.533}^{+5.525}$ & $55813.763_{-0.367}^{+0.464}$ & $0.613_{-0.301}^{+0.344}$ & O-\\
... & ... & ... & ... & ... & ... & ... & ... & ... \\
\hline
    \end{tabular}
    \label{tab:event_inference}
\end{table*}
\endgroup


\bsp	
\label{lastpage}
\end{document}